\documentclass{aa}

\usepackage{graphicx}
\usepackage{dblfloatfix} 
\usepackage{multirow}
\usepackage{txfonts}
\usepackage{natbib}
\usepackage{hyperref}
\usepackage{hhline}
\usepackage{float}
\usepackage{lscape}
\usepackage{xcolor}
\hypersetup{
    colorlinks=true,
    linkcolor=magenta,
    citecolor=blue,
    urlcolor=magenta,
    }

\newcommand{\rvir}{R_{vir}}

\newcommand{\vth}{v_{200}}

\newcommand{\flya}{F_{Ly\alpha}}
\newcommand{\laf}{\text{Lyman-}\alpha}
\newcommand{\vsbla}{v_{\text{SBLA}}}
\newcommand{\ptot}{P_{tot}}
\newcommand{\pmh}{P(M_h)}
\newcommand{\kms}{\,\,{\rm km}\,{\rm s}^{-1}}
\newcommand{\cf}{C_{Frac}}

\defcitealias{pieri10}{P10}
\defcitealias{pieri14}{P14}
\defcitealias{morrison24}{M24}

\bibpunct{(}{)}{;}{a}{}{,}

\begin{document}

\title{Finding Halos in the Lyman-$\alpha$ Forest}
\subtitle{I. Introducing a hierarchical absorption framework}
\author{Duarte Muñoz Santos\inst{1} \and Matthew M. Pieri \inst{1} \and Dylan Nelson\inst{2} \and Teng Hu\inst{1} \and Simon Weng\inst{1} \and Manuel F. Ruiz-Herrera Bernal\inst{3,4}}

\institute{Aix Marseille Univ, CNRS, CNES, LAM, Marseille, France \\ E-mail: \href{mailto:duarte.santos@lam.fr}{\texttt{duarte.santos@lam.fr}} \and Universität Heidelberg, Zentrum für Astronomie, ITA, Albert-Ueberle-Str. 2, 69120 Heidelberg, Germany \and CIEMAT, Avenida Complutense 40, E-28040 Madrid, Spain \and Facultad de F\'isicas, Universidad de Sevilla, Avda. Reina Mercedes s/n, Campus de Reina Mercedes, 41012 Sevilla, Spain}

\date{Received ?? \ Accepted ??}

\abstract{It has been demonstrated  that one can track down galaxies in absorption `hidden' in the $\laf$ forest through the use of `strong, blended $\laf$' (or SBLA) absorption. Specifically a series of publications studied SBLA absorption systems with $\laf$ flux transmission, $\flya < 0.25$ on scales of $138\kms$ in the Sloan Digital Sky Survey (SDSS).
In order to better understand the connection between halos and these SBLAs, we make use of several million synthetic absorption spectra from the TNG50 cosmological simulation, at $z = 2$ and $z = 3$.
We explore spectra with SDSS-like resolution in order to understand the nature of SBLAs as defined thus far, as well as with high resolution (or `resolved') spectra to generalise and optimise SBLAs as halo finders.
For the SDSS SBLAs, we find that up to 67\% of these absorption systems reside in halos, where the stronger the absorption and the lower the redshift, the higher the probability.
We also manage to recover a mean halo mass of $10^{11.78} \: M_{\odot}$, broadly in line with what is measured in observations. For the resolved SBLAs, we expand on the previous definition and allow the SBLA spectral size to vary between $54 \kms$ and $483 \kms$. We find that the largest absorbers have the highest probability of finding halos (at 72\%). When applying a hierarchical framework, where we allow the largest SBLAs to consume the smaller ones, we find that the halo mass distributions for each SBLA spectral size becomes narrower with respect to the non-hierarchical case. We are also able to probe halo masses from $M_h \approx 10^{9.5} M_{\odot}$ (for $100 \kms$ SBLAs) to $M_h \approx 10^{11.5} M_{\odot}$ (for $450 \kms$ SBLAs).
With these results, we show that we are able to transform the $\laf$ forest into a halo finding machine for not only identifying CGM regions, but also estimating their host halo masses.}

\keywords{Galaxies: evolution - Galaxies: intergalactic medium - Quasars: absorption lines - Cosmology: theory - Techniques: spectroscopic - Methods: numerical}

\maketitle

\section{Introduction}\label{section:intro}

Ever since the `Great Debate' \citep{Shapley21,Hoskin76}, the existence of `island universes' has shaped our view of the Universe. Now, a little over a century later, we have come to a new understanding that these objects, which we now call galaxies, are pieces of the puzzle that help us understand the history of the Universe. By tracing their formation and evolution, we can better understand how we have reached our current epoch.

Galaxies are not isolated systems. They are a part of the tangled cosmic web of the Universe, interacting not only with each other (e.g. Messier 51, \citealt{Toomre72}), but also with the medium around them (e.g. \citealt{bergeron86}). Understanding the relationship between the intergalactic medium (IGM) and galaxies allows us to paint a picture of how galaxies form and evolve. The IGM contributes pristine gas for star-formation, while galaxies eject metals into the IGM via outflows driven by active galactic nuclei (AGN) or stellar feedback \citep{tinsley80,lilly13,dekel14}.

After the discovery of quasars (or QSOs) by \cite{Schimdt63}, the analysis of the restframe ultraviolet spectra of these objects revealed a trough in absorption \citep{gunnpeterson65}. However, with higher resolution spectra, a `forest' of absorption blueward of the $\laf$ emission line was revealed, first described as a set of discrete clouds in the IGM \citep{BahcallSpitzer69}, but nowadays as a fluctuating density field of neutral hydrogen gas, or the fluctuating Gunn-Peterson approximation (e.g. \citealt{croft98,rauch98,weinberg03}). This allows us to, for example, trace the properties of the medium between galaxies such as density, temperature and metallicity (e.g. \citealt{Hu25}).

The region of the IGM around galaxies has come to be known as the  circumgalactic medium (CGM), in order to isolate the interface region in which galaxies and the wider IGM interact. Specifically the CGM is typically defined as the dark matter halo's gravitationally bound zone as estimated by its virial radius and virial velocity. Hereafter, when we use the term `IGM' we refer to the portion of the wider IGM that excludes CGM regions, for simplicity. The CGM is one of the main baryonic reservoirs of the Universe (alongside the IGM), so understanding its relationship with galaxies and the IGM, as well as how it has changed and evolved with time is of crucial importance. There are many methods to observe and analyse this medium, ranging from direct emission line maps (e.g. \citealt{Lusso19,Martin23,Tonotti25}), to `down the barrel' spectroscopy (e.g. \citealt{Henry15,Heckman15}), but by far the most common is through absorption-line spectroscopy.

Although the diffuse IGM is traced by the $\laf$ forest, it is not exclusively created by diffuse gas. When lines-of-sight pass through the CGM, $\laf$ forest absorption may also be generated \citep{lanzetta95,barcons95,chen98}. In the simulations studied here, we find that 1.2\% of the $\laf$ forest is associated with the CGM, and while this may seem like a small number, if we consider large volume surveys such as the Dark Energy Spectroscopic Instrument (DESI, \citealt{desi_inst}), 
where $\sim 1$ million $\laf$ quasars have been observed, this provides us with a redshift path of $\Delta z \gtrsim 1\,000$ to study, with more to come from surveys such as WEAVE-QSO \citep{weaveqso_survey}.

The standard way to trace galaxies in quasar absorption spectra is through high hydrogen column density systems, such as Damped $\laf$ (DLA) systems ($N_{HI} \gtrsim 10^{20} \: \text{cm}^{-2}$, e.g. \citealt{wolfe05,chen05,peroux22,weng23,Nunez24}), 
or Lyman Limit Systems (LLS, $ 10^{17} \: \text{cm}^{-2}\lesssim N_{HI} \lesssim 10^{20}\:\text{cm}^{-2}$, e.g. \citealt{fumagalli13, lehner13}). In reality a wider range of gas conditions, and therefore column densities, are present in the CGM. Simulations such as \cite{hummels19} and \cite{Oppenheimer20} show that a large fraction of the CGM ($\gtrsim 90\%$) is made up of lower column density systems ($10^{12} \: \text{cm}^{-2} \lesssim N_{HI} \lesssim 10^{17} \: \text{cm}^{-2}$). Furthermore, when comparing the incidence rate of halos ($dn/dz = 9.1$ at $z=2$, and $dn/dz = 6.7$ at $z=3$, as estimated in the simulation studied here) with those of DLAs and LLS ($dn/dz=0.06$ and $dn/dz=1.21$, respectively; \citealt{prochaska05,fumagalli13}) we can see that we are limiting ourselves to a biased subset of CGM systems and approximately an order of magnitude fewer systems than is traced in absorption.

Finding galaxies through lower column density systems is a challenge, as distinguishing the portions of the $\laf$ forest that trace the CGM instead of the IGM is a non-trivial matter. In \citealt{pieri10} \citepalias[hereafter][]{pieri10}, using data from the Sloan Digital Sky Survey (SDSS, \citealt{sdss_technical1,sdss_technical2,sdss_technical3}), the authors serendipitously found that absorbers with $\laf$ flux transmission below 25\% and on 138$\kms$ scales in the forest are typically associated with optically thin CGM gas. \citealt{pieri14} \citepalias[hereafter][]{pieri14} studied these systems in greater detail and also explored a sample of quasars with deep data with projected close-by Lyman break galaxies (within 300 proper kpc). They studied strong absorption complexes of varying strengths and found that over half of the time, a galaxy was present (when completeness corrected), when the flux transmission was below 25\%.

In \citealt{morrison24} \citepalias[hereafter][]{morrison24} these objects were given the name of `strong, blended Lyman-$\alpha$' (SBLA), because they arise from $\laf$ absorption complexes that lead to low transmission on an extended scale. \citetalias{pieri10}, \citetalias{pieri14}, and \citetalias{morrison24} studied samples with a scale set by $138 \kms$ bins plus instrumental broadening, but \citetalias{morrison24} recognised that this choice was somewhat arbitrary (which is natural since the discovery that they probe the CGM was serendipitous). They defined SBLAs more generically as strong absorption structures blended together on some extended scale leaving open the possibility that SBLA selection can be optimised and that the significance of the absorption scale can be explored. It is that wider exploration that motivates this work. The authors stacked the restframe spectra of this sample (with a median $z = 2.7$) which provided insights into the density and metallicity of different phases of the CGM. Particularly, studying the low-ionisation species appears to allow for the detection of parsec-scale clumping (driven by the observations of the \ion{Si}{II} and \ion{C}{II} absorbing population) and even sub-parsec scale clumping (driven by the \ion{O}{I} absorbing population). Furthermore, by cross-correlating their SBLA sample with the $\laf$ forest in all of their SDSS data, they measured the large-scale structure bias and were able infer the halo mass associated with these CGM regions, which has a range of values between $10^{12} M_{\odot}$ and $10^{12.3} M_{\odot}$. 

In effect, \citetalias{pieri10}, \citetalias{pieri14}, and \citetalias{morrison24} advocate throwing off the limitation that physical systems be found by (and associated with) a single line. Instead they argue that CGM regions as a whole may imprint themselves as an extended absorption complex like a form of CGM `fingerprint'.

This is not the only method to detect the CGM through absorbing structures  -- another example comes from the MApping the Most Massive Overdensity Through Hydrogen (MAMMOTH, \citealt{mammothI,mammothII}) project, where strong absorption over large scales ($15-30 \: \AA$ at $z=2.5$, or about $1025-2143 \: \kms$, in $15-30 \: h^{-1}$ Mpc scales) is associated with $\laf$ emitters. However, these objects are rare, and the goal of the authors was to find the highest mass tail of the halo mass function.

SBLAs, then, are a useful alternative when compared with their higher column density counterparts because they are more numerous (main analysis sample of \citetalias{morrison24} has an incidence rate of $dn/dz = 12.4$), likely able to trace many of the halos that DLAs and LLS cannot; probe different regions, phases, and structures of the CGM; and provide insights into the structure of the medium. 
In order to maximise the utility of SBLAs to the community, questions must be addressed that have been partially or fully unanswered thus far. For example `how often do these absorption complexes trace the CGM?', `what halos do they trace?', `what is the covering fraction of the CGM?', and `how are the answers to the previous questions affected by selection details?'. Overall, the available information is severely limited by the small samples of galaxies currently available above $z=2$ with associated CGM information.

In this first paper, we wish to better understand the connection between halos and SBLAs, at $z > 2$. To do so, we use the TNG50 simulation \citep{tng50base_1,tng50base_2} as a model universe to build greater insight into these objects. In \citetalias{morrison24}, the authors examined samples with SDSS resolution, but stressed that the SBLA framework was not limited to these specific samples. We take up this challenge by not only simulating their SBLA samples, but also by considering an alternative general approach to use this framework to explore a wider range of halos. The following paper in this series (Hu et al. in prep.) will tackle the effects of instrumental broadening and noise in the SBLA selection function.

This paper is structured as follows: in Sect. \ref{sec:hydro spectra} we briefly describe the TNG50 simulation, as well as what data we use from it; in Sect. \ref{sec:sbla finding} we describe the selection process of SBLAs, and how we associate them with galaxy halos; in Sect. \ref{sec:SDSS-like} we show the results of SBLAs that have the same properties as those of \citetalias{morrison24}; in Sect. \ref{sec:variable size} we explore SBLAs with a range of absorption blend sizes and study the properties of the halos found; in Sect. \ref{subsec:hierarchy} we introduce a hierarchical approach (which preserves only the larger SBLAs when SBLAs on different scales overlap), and study the joint halo properties of the SBLA hierarchy as a strategy for building halo candidate lists; in Sect. \ref{sec:discussion} we discuss our results, and in Sect. \ref{sec:conclusion} we provide our conclusions.

\begin{figure*}[h]
    \centering
    \includegraphics[width=0.9\linewidth]{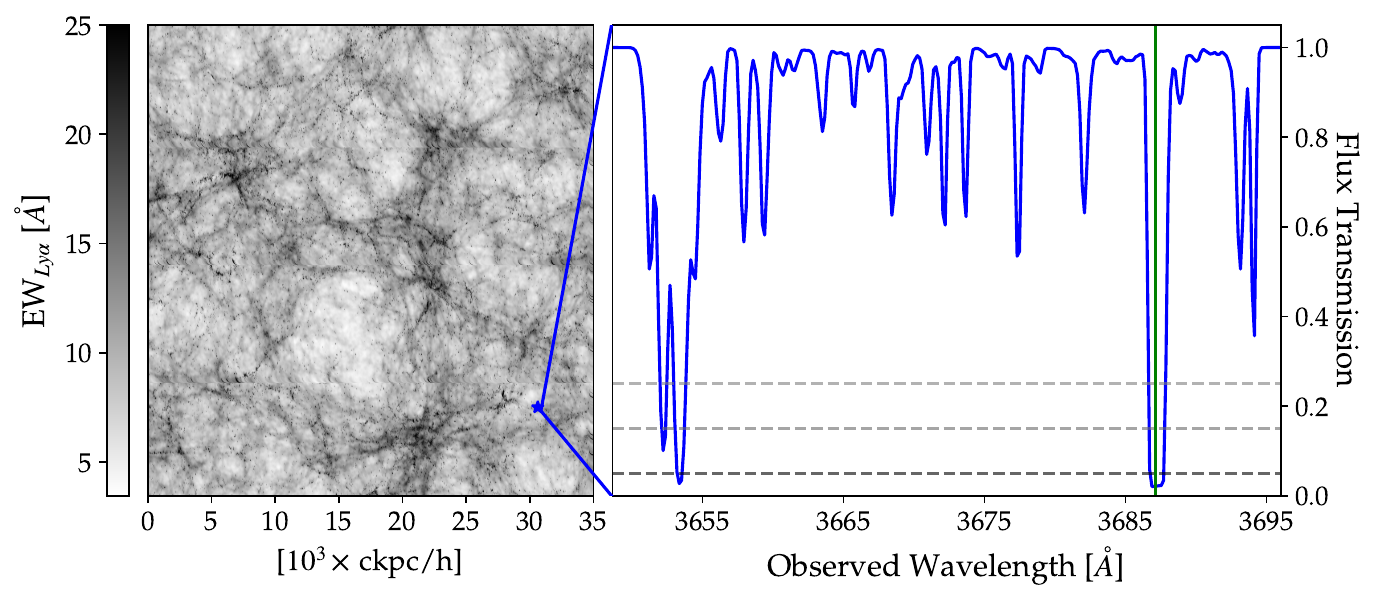}
    \caption{A randomly-selected resolved spectrum (right) from the TNG50 simulation box (left). The spectrum has the `resolved' resolution at $z=2$, whose position is denoted by the blue star on the left. Only the Lyman-$\alpha$ transitions are shown. In the box coordinates, the x- and y-axis have the same dimensions, and this data is taken from the evenly sampled lines-of-sight list, with a spacing of 35 ckpc/h. The colourbar represents the integrated equivalent width (EW) of the Lyman-$\alpha$ transition for each spectrum. The green vertical line marks the position of a halo in the spectrum. The darkening horizontal dashed lines represent the three different samples we consider for our work: $\flya < 0.25$, $\flya < 0.15$ and $\flya < 0.05$.}
    \label{fig:randspec sdss}
\end{figure*}

\section{Hydrodynamic Simulations and Synthetic Spectra}\label{sec:hydro spectra}

The relationship between the CGM and SBLAs has been explored in the literature \citepalias{pieri10,pieri14,morrison24}, and here we wish to explore it further by appealing to simulations. Our goals are twofold: to produce SDSS-like spectra to explore the above works directly, and to produce resolved $\laf$ forest spectra to developed a more general and optimised framework for SBLA sample building.

This work uses TNG50\footnote{\hyperlink{https://www.tng-project.org/data/}{https://www.tng-project.org/data/}} \citep{tng50base_1,tng50base_2}, the highest-resolution version of the IllustrisTNG suite of large-scale cosmological magneto-hydrodynamical simulations \citep{Marinacci2018, Naiman2018, Nelson2018, Pillepich2018b, Springel2018}. IllustrisTNG was built upon the foundation of the original Illustris simulation \citep{Genel2014, Vogelsberger2014a, Vogelsberger2014b, Sijacki2015}, now incorporating magnetic fields \citep{Pakmor2014} and with changes on the original galaxy formation model \citep{Vogelsberger2013, Torrey2014} through the introduction of updated feedback mechanisms \citep{Weinberger2017, Pillepich2018a}. As the name implies, the TNG50 simulation spans roughly $50$ comoving Mpc$^3$ (cMpc$^3$), or 35 (cMpc/h)$^3$, with $H_0 = 67.74 \: \text{km} \: \text{s}^{-1} \: \text{Mpc}^{-1}$ and a cosmology of $\Omega_0 = 0.3089$ and $\Omega_{\Lambda} = 0.6911$, and has a resolution comprising $2160^3$ elements, attaining a baryonic (dark matter) mass resolution of $8.5 \times 10^4$ M$_\odot$ ($4.5 \times 10^5$ M$_\odot$). The combination of a substantial volume and high spatial resolution makes it particularly well-suited for examining the CGM of galaxies. We considered the simulation box at two redshifts: $z = 2$ and $z = 3$, in order to study how the absorber-halo relationship is affected by different epochs of the Universe.

We made use of 4 million parallel line-of-sight noiseless synthetic transmission spectra that are randomly spread across the TNG50 simulation box and include only the Lyman series transition lines (based on the work presented in \citealt{nelson25}). In the Lyman-$\alpha$ transition, this equates to (generally) a minimum wavelength of $3645 \: \AA$ and $4870 \: \AA$, and a maximum wavelength of $3695 \: \AA$ and $4930 \: \AA$, for $z=2$ and $z=3$, respectively.\footnote{This translates into wavelength intervals of about $50 \: \AA$ per spectrum for $z=2$ and about $60 \: \AA$ per spectrum for $z=3$.} Hereafter when we use the term `lines-of-sight' it should be noted that we refer to these short stretches of absorption spectrum and not the hundreds of $\aa$ngstr\"oms typically observed in Lyman-$\alpha$ forest quasars. In order to calculate the ionisation state of hydrogen, the ultraviolet background from \cite{faucher-guigere09} was used; the HI density was estimated based on the neutral hydrogen abundance computed during the simulation run, which uses on-the-fly self-shielding based on the \cite{rahmati13} model -- this is re-computed for dense, star-forming gas in post-processing; and the molecular H$_2$ was removed using the pressure-based model of \citet{blitz06}.

We generate a sample of synthetic spectra with KECK/HIRES-like instrumental broadening ($4.5\kms$) for realism, but since the broadening is a small correction to the inherent line widths we refer to this sample hereafter as `resolved spectra'.
See Fig. \ref{fig:randspec sdss} for an example of a resolved spectrum, with a halo marked. This sample is explored in Sect.~\ref{sec:variable size} using  varying SBLA sizes for a more general SBLA framework.

We extracted properties from the halos present in the simulation, namely: the three dimensional position of the centre of mass of halos; $R_{vir}$ is the virial radius of halos, which we define as the comoving radius of a sphere centred on a halo where the mean overdensity is $\rho/\bar{\rho} = 200$, where $\bar{\rho}$ is the mean density of the Universe; $M_h$ is the mass of each halo; and $v_{pec}$ is the peculiar velocity of the halo. We also considered the theoretical value for the maximum circular velocity of each halo which we define as:
\begin{equation}
    v_{200} = \sqrt{\frac{G M_{h}}{R_{vir}}}
    \label{eq:v200}
\end{equation}
where $G$ represents the gravitational constant, $M_{h}$ represents the mass of the halo, and $R_{vir}$ its associated virial radius.

In order to make the statistics of our synthetic absorption spectra as realistic as possible, we post-process them as follows. Firstly, we rescale their mean flux transmission at each redshift to obtain $\bar{F} \approx 0.86$ at $z=2$ and $\bar{F} \approx 0.68$ at $z=3$ in line with \citet{fauchergiguere08} and \citet{mdonald2000}. We did this by rescaling the optical depth using a constant derived from the ratio of the target mean and the measured sample value. Secondly, we remove all the lines-of-sight that include Damped Lyman-$\alpha$ (DLA) or sub-DLA absorbers, so they do not contaminate our sample of SBLAs. To do this, we conservatively selected the lines-of-sight that have integrated hydrogen column density $N_{HI} > 10^{19} \: \text{cm}^{-2}$, accounting for $81\,733$ lines-of-sight at $z=2$ and $204\,815$ lines-of sight at $z=3$ -- this is equivalent to $2\%$ and $6\%$, respectively, of the total available lines-of-sight. We remove these systems because, while they may be used to identify CGM systems, they are detected due to gas cloud properties rather than the global halo properties that given rise to line blending. Furthermore, while these systems are rare, they are also well-studied, and we wish to conservatively study the value added by the SBLA framework that we explore.

With these adjustments, we then smoothed and rebinned the resolved spectra so they had the same resolution as SDSS-BOSS spectra ($69 \kms$), providing us with a direct comparison of previous observational work \citepalias{pieri10,pieri14,morrison24}. 

\section{General SBLA definition and Halo Matching}\label{sec:sbla finding}

In the previous section we provided details about the TNG50 simulation and the synthetic spectra we will be using in this work. Here, we explain how we select SBLA samples, and how to associate them with halos. 

\begin{figure}[h]
    \centering
    \includegraphics[width=0.95\linewidth]{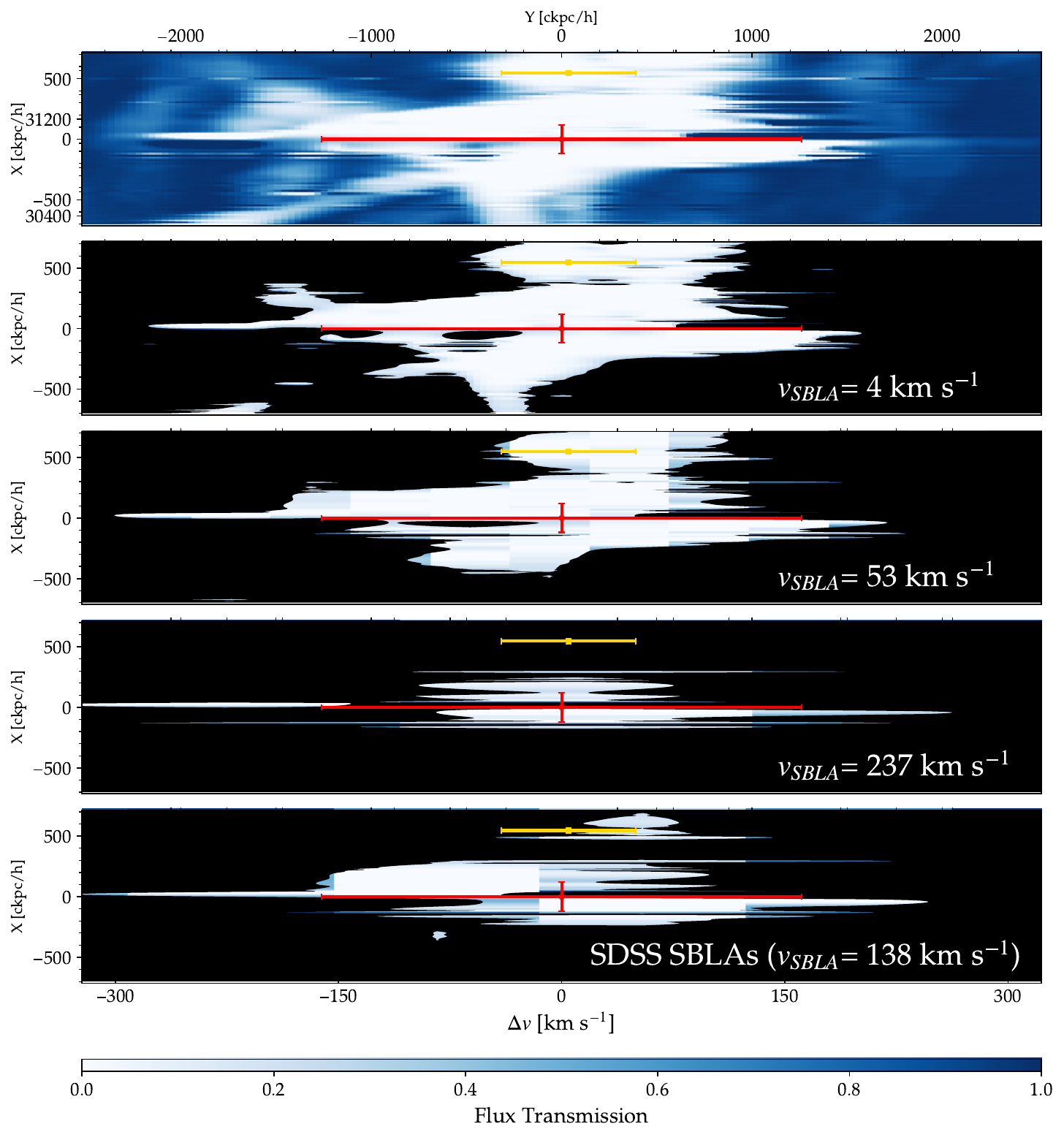}
    \caption{Representation of two halos in physical and line-of-sight dimensions. On the y-axis and the top x-axis, we have the physical dimensions, and on the bottom x-axis, we have the the line-of-sight direction. The red lines represent a halo of mass $10^{11.54} \: M_{\odot}$ and the yellow lines represent a halo of mass $10^{10.04} \: M_{\odot}$. The top panel represents the $\laf$ flux transmission without restrictions, as shown on the bottom colourbar. The three panels below represent a limit on the flux transmission of 25\%, with the binning of the spectrum, $\vsbla$, increasing from top to bottom: 4$\kms$, 53$\kms$ and 237$\kms$. The bottom panel represents an SDSS-like SBLA, with $138 \kms$.}
    \label{fig:halo example}
\end{figure}

As mentioned in Sect. \ref{section:intro}, SBLAs are defined as extended structures with strong absorption in the $\laf$ forest. We keep \citetalias{pieri14} and \citetalias{morrison24} limits of flux transmission as `strong' (that being, flux transmission below 0.25, 0.15 and 0.05), but allow the scale of the structures to be changed, so we can explore different types of SBLA samples. In Fig. \ref{fig:halo example} we can see a series of nearby transmission spectra that probe a region around two halos: one of mass $10^{11.54} \: M_{\odot}$ (represented in red) and another of mass $10^{10.04} \: M_{\odot}$ (represented in yellow). These halos are shown in one spatial direction and in the line-of-sight direction for the native spectral resolution, including different spectral sizes and SDSS-like SBLAs. In binned cases we mask everything except flux transmission below 25\% to illustrate the SBLA selection on different scales.

In Sect. \ref{sec:SDSS-like} we will explore relationship between SDSS-like SBLA and the halos they are associated with, and show how to compare these results with observations. In Sect. \ref{sec:variable size} we will explore a generalisation of SBLA sample-building and predict the range of halos one might expect to find.

We will now define the framework we use for spatially and spectrally associating SBLAs with halos. We considered three criteria to achieve this: 1) halos must have masses above $10^9 \: M_{\odot}$, 2) the line-of-sight hosting the SBLA should pass within the $\rvir$ of the halo, and 3) the spectral range associated with the SBLA must overlap with the spectral range in the Lyman-$\alpha$ forest described by the redshift and the circular velocity of the halo.

For point 2) to be true, we apply a simple condition, where he have that:
\begin{equation}
    (X_{\text{SBLA}} - X_{\text{Halo}})^2 + (Y_{\text{SBLA}} - Y_{\text{Halo}})^2 \leqslant \rvir^2
        \label{eq:rvir}
\end{equation}
where $X_{\text{SBLA}}$ and $Y_{\text{SBLA}}$ represent the positions of the lines-of-sight that hold SBLAs and $X_{\text{Halo}}$ and $Y_{\text{Halo}}$ represent the positions of the halo, in the simulation box.

For point 3) to be true, we need to ensure that the redshift difference between the halo and the SBLA, expressed as a velocity, is less than the halo circular velocity. In other words: 
\begin{equation}
    \Delta v =  \frac{|z_{\text{SBLA}} - z_{\text{Halo}}|}{1 + z_{\text{Halo}}} \; c \leqslant v_{200}
    \label{eq:deltav}
\end{equation}
where $c$ represents the speed of light in$\kms$, $z_{\text{SBLA}}$ represents the redshift of the SBLA, $z_{\text{Halo}}$ represents the redshift of the halo and $v_{200}$ the halo circular velocity defined in Eq. \ref{eq:v200}.

\section{SDSS-like SBLAs} \label{sec:SDSS-like}

\begin{figure*}[h]
    \centering
    \includegraphics[width=0.85\linewidth]{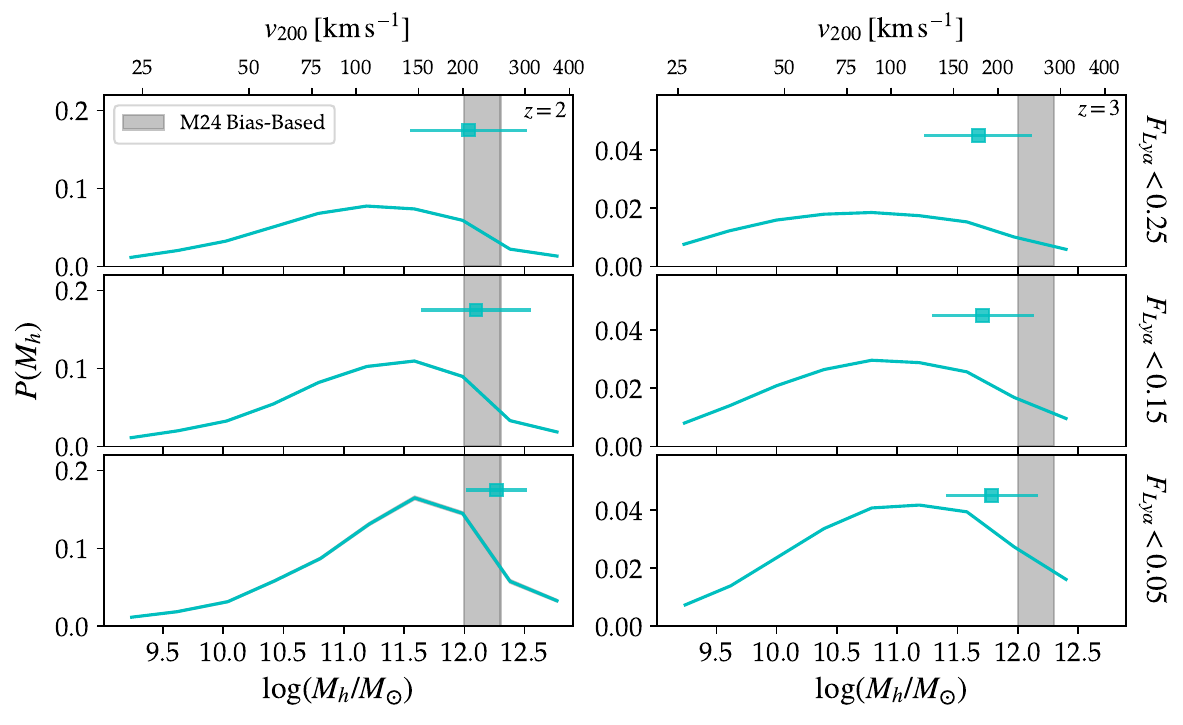}
    \caption{$\pmh$ of the SDSS SBLA samples ($\vsbla = 138\kms$) spectra. $z=2$ results are on the left panels and $z=3$ results are on the right. The top axis represent the equivalent halo circular velocity for each halo mass, according to a linear fit we performed. The grey shaded area represents the halo mass estimate from SBLA bias in \citetalias{morrison24}. The square points and error bars represent, respectively, the mean halo mass traced by that SBLA sample and the $1\sigma$ deviation. Each row represents, from top to bottom, a different flux limit: $\flya < 0.25$, $\flya < 0.15$ and $\flya < 0.05$.}
    \label{fig:psbla sdss random}
\end{figure*}

In this section, we explore the SBLA sample as found in the literature. \citetalias{pieri14} and \citetalias{morrison24} rebinned SDSS spectra with $\text{S/N/\AA} >3$ by 2 pixels (corresponding to a spectral size of $\vsbla = 138 \kms$), and studied three SBLA samples: FS2, with $0.15 < \flya < 0.25$; FS1, with $0.05 < \flya < 0.15$; and FS0, with $-0.05 < \flya < 0.05$. It is important to note that observing noise is a key contributor to those samples. Specifically \citetalias{pieri14} found that, based on a comparison to Lyman break galaxies, the goal should be to obtain a true (noiseless) flux transmission $\flya < 0.25$, and they achieved this for FS0, FS1 and FS2 with 90\%, 79\% and 49\% purity respectively (when conservatively assuming that all the spectra have $\text{S/N/\AA} = 3$). We will explore this claim by comparing with SDSS-like spectra generated from the TNG50 simulation. To do this, we will calculate how many SBLAs reside in galaxy halos, and how many SBLAs cover the virial radii of these halos.

Firstly, we must obtain our SBLA samples. To do so, we rebinned our SDSS-like spectra by 2 pixels, which translates into an SBLA spectral size of $\vsbla = 138 \kms$. We use the standard sample of SBLAs stated in \citetalias{morrison24}, where the noiseless flux transmission must be $\flya < 0.25$. We explore SBLAs in greater detail by introducing more selective samples, limited to stronger absorption ($\flya < 0.15$ and $\flya < 0.05$).

As can be seen on Table \ref{tab:sblacounts sdss in}, depending on the flux limits, we find between $10^4$ to nearly $10^6$ SBLAs at $z=2$, and $10^6$ to nearly $10^7$ SBLAs at $z=3$, with the highest flux limit ($\flya < 0.25$) possessing the largest number of these objects. Overall, we find that between $1\% - 18\%$ of lines of sight through TNG50 have SBLAs at $z=2$, while at $z=3$ we have between $13\% - 69\%$. This does not take into the account the spectra with DLAs. Furthermore, we calculated the incidence rates of SBLAs, which we explain with more detail in Appendix \ref{appendix:sbla props}.

\begin{table}[h]
\caption{Total number of SBLAs, $n_{\text{SBLA}}$, the respective probability of finding them in halos, $\ptot$, and the incidence rates, in units of $dn/dz$.}
\centering
\label{tab:sblacounts sdss in}
\setlength{\tabcolsep}{6pt}
\renewcommand{\arraystretch}{1.2}
\begin{tabular}{ccccc}
\multicolumn{5}{c}{$\vsbla= 138\ \text{km}  \: \text{s}^{-1}$ at SDSS resolution} \\ \hline \hline
\multicolumn{1}{l}{} & \multicolumn{1}{l}{} & $\flya <0.25$ & $\flya <0.15$ & $\flya <0.05$ \\ \hline
\multicolumn{1}{c|}{\multirow{3}{*}{$z=2$}} & \multicolumn{1}{c|}{$n_{\text{SBLA}}$} & $984\,941$ & 352$\,$396 & 41$\,$021 \\
\multicolumn{1}{c|}{} & \multicolumn{1}{c|}{$\ptot$} & 0.40 & 0.51 & 0.67 \\
\multicolumn{1}{c|}{} & \multicolumn{1}{c|}{$dn/dz$} & 6.11 & 2.19 & 0.25 \\ \hline
\multicolumn{1}{c|}{\multirow{3}{*}{$z=3$}} & \multicolumn{1}{c|}{$n_{\text{SBLA}}$} & 7$\,$585$\,$471 & 3$\,$836$\,$944 & 1$\,$568$\,$869 \\
\multicolumn{1}{c|}{} & \multicolumn{1}{c|}{$\ptot$} & 0.11 & 0.17 & 0.23 \\
\multicolumn{1}{c|}{} & \multicolumn{1}{l|}{$dn/dz$} & 40.5 & 20.5 & 8.38 \\ \hline
\end{tabular}
\end{table}

\subsection{Total Probability for SDSS SBLAs}\label{subsec:SDSS probs}

Following the procedure laid out in Sect. \ref{sec:sbla finding} to associate SBLAs with halos, we can then define the total probability of finding an SBLA inside a halo as $\ptot$, seen below.

\begin{equation}
    \ptot = \frac{\text{Number of SBLAs in halos} }{\text{Total number of SBLAs} }
    \label{eq: psbla}
\end{equation}

SBLAs with $\vsbla = 138 \kms$ are found inside halos with $\ptot$ values that range between $11-67\%$, depending on how restrictive we are with the flux limits and the redshift considered (see Table \ref{tab:sblacounts sdss in} for more details).

The main SBLA sample studied in \citetalias{morrison24} is the FS0 sample, which has noise-in $\laf$ flux transmission below $5\%$. However, referring to Fig. 3 of \citetalias{pieri14} for $\text{S/N/\AA} = 3$ (which is the minimum signal-to-noise of that work), SBLAs selected in the FS0 most closely resemble our noiseless $\flya < 0.15$ sample. 

\begin{figure*}[h]
    \centering
    \includegraphics[width=0.85\linewidth]{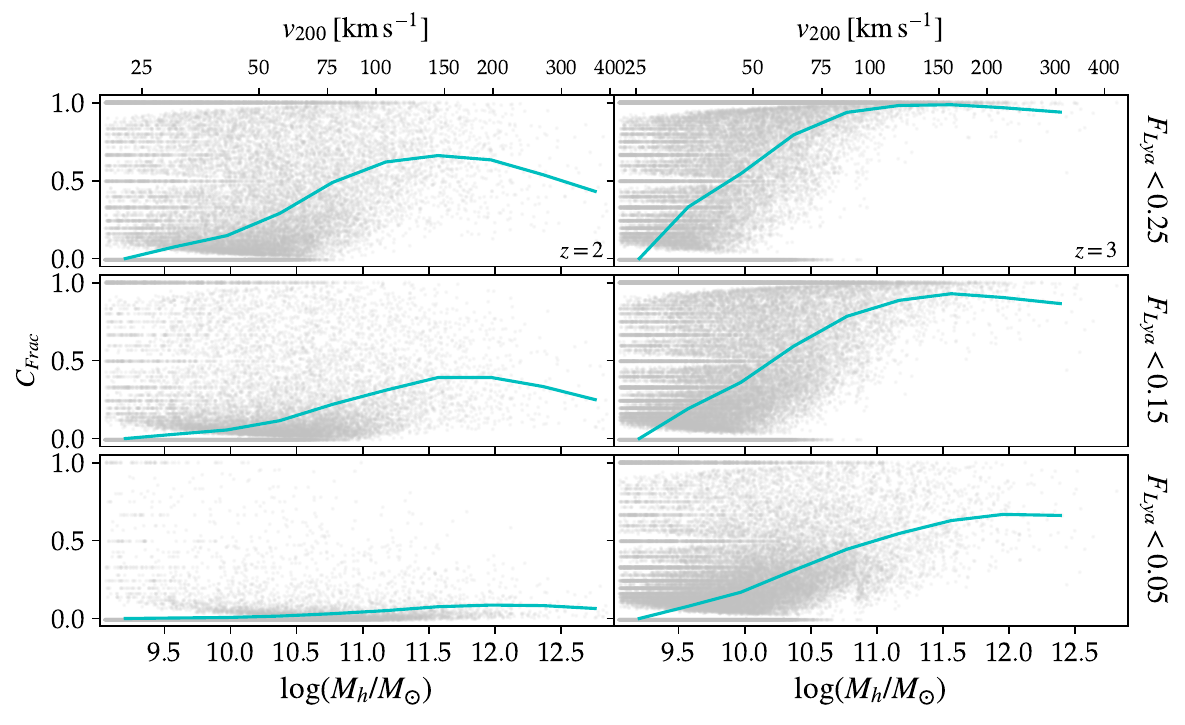}
    \caption{Covering fraction of the SDSS SBLA samples ($\vsbla = 138\kms$) spectra, divided in mass bins. $z=2$ results are on the left panels and $z=3$ results are on the right. The grey points represent the covering fraction for each individual halo, while the solid cyan line represents the mean covering fraction for each mass bin mentioned in the text. All remaining elements are the same as in Fig. \ref{fig:psbla sdss random}. The horizontal `bars' are collections of points with the same percentage -- due to the resolution of the simulation, we find that many of the lower mass halos have covering fractions that are discrete and not continuous.}
    \label{fig:cov fraction sdss random}
\end{figure*}

At $z=2$ ($z=3$), we find that $\ptot$ is 51\% (17\%)  for the $\flya < 0.15$ sample. According to \citetalias{pieri14}, over half the SBLAs selected in these samples should be associated with Lyman break galaxies (including completeness corrections). \citetalias{morrison24}  went a step further and forward modelled the metal populations associated with FS0 (with the same signal-to-noise requirement). The modelling in that work (and others) show that it is implausible for any low ionisation species (such as \ion{Si}{ii}, \ion{C}{ii} and \ion{O}{i}) to be detected in the IGM. The stronger populations of these species were modelled to belong to the CGM in 20\% to 40\% of cases. This is a conservative assessment of CGM incidence because it is unlikely that the rather extreme conditions required to detect these species are present every time a line-of-sight passes through the CGM. Instead, for a more accurate assessment, one may refer to the size of the strong populations of intermediate (e.g. \ion{Si}{iii}) or high (e.g. \ion{C}{iv}) ions, which indicate CGM incidence of between 35\% and 60\%. Our $z=2$ results appear to be broadly consistent with these observational analyses but our $z=3$ results do not. \citetalias{pieri14} and \citetalias{morrison24} study SBLAs with $2.4<z<3.1$ with a median $z=2.7$ which bisects our two redshift boxes. Superficially one might expect $z=3$ (as the closer redshift) to be more relevant, but substantial and non-linear evolution occurs from $z=3$ to $z=2$ so the correct choice is non-trivial. 

Overall the results are encouraging and support the claims of \citetalias{pieri14} and \citetalias{morrison24}. As demonstrated in \citetalias{morrison24} a dilute population of hundreds of thousands of previously unidentified halos mixed in with IGM contamination can be exploited with careful statistical methods. 

Finally we note that our $\flya < 0.05$ sample shows $\sim 15\%$ higher purity than the $\flya < 0.15$. The above observational works were not able to exploit this potentially improved CGM-finding fidelity given their use of data with signal-to-noise as low as $\text{S/N/\AA} = 3$. Our results indicate that significant improvements can be made with data of sufficiently high signal-to-noise to reliably recover this lower flux transmission.

\subsection{Halo Mass Dependence for SDSS SBLAs}\label{subsec: SDSS halo mass}

$\ptot$ is a useful metric for how SBLAs are tracing halos (and so, CGM regions) overall, but it is ambiguous as it neglects information about the mass of the halos found. Here we proceed to explore this mass dependence.

In order to study the relationship between SBLAs and the masses of halos, we separated halos into mass bins of width $10^{0.4} \: M_{\odot}$. We do this to select at least 15 halos per bin. With this criteria, we are able to study halos masses from $10^9 \: M_{\odot}$ to $10^{13} \: M_{\odot}$ at $z=2$, and at $z=3$ we are able to study halo masses from $10^9 \: M_{\odot}$ to $10^{12.6} \: M_{\odot}$. In Table \ref{tab: halo number per mass bin} we show the distribution of the number of halos per bin.

\begin{table}[h]
\caption{Number of halos, $n_h$, per mass bin, for each redshift.}
\label{tab: halo number per mass bin}
\centering
\begin{tabular}{l|ll}
&  \multicolumn{1}{c}{$z=2$}& \multicolumn{1}{c}{$z=3$} \\ \hline
Mass bin {[}$\log(M_h/M_{\odot})${]} & \multicolumn{1}{c}{$n_{h}$} & \multicolumn{1}{c}{$n_{h}$}  \\ \hline
{[}9, 9.4)     & $68\,482$ & $70\,042$ \\
{[}9.4, 9.8)   & $29\,382$ & $29\,485$ \\
{[}9.8, 10.2)  & $12\,863$ & $12\,655$ \\
{[}10.2, 10.6) & $5\,562$  & $5\,260$  \\
{[}10.6, 11)   & $2\,412$  & $2\,085$  \\
{[}11, 11.4)   & $1\,043$  & 775   \\
{[}11.4, 11.8) & 444   & 285   \\
{[}11.8, 12.2) & 172   & 85    \\
{[}12.2, 12.6) & 40    & 23    \\
{[}12.6,  13)  & 17    & ---     \\
$> 13$  & ---  & --- \\ \hline
\end{tabular}
\tablefoot{The horizontal bars represent mass bins we did not use in our analysis due to the limited number of halos present at those mass ranges.}
\end{table}

We then found how many SBLAs are associated with halos in a given mass bin following the procedure set out in Sect. \ref{sec:sbla finding}, and divided this number by the total number of SBLAs found, as per Eq. \ref{eq: psbla} -- we called this new statistic $\pmh$.

Note that each mass bin probability is calculated independently of the others and that integrating $\pmh$ over the mass range to be studied does not yield $\ptot$, but an overestimation of it. SBLAs can be associated with more than one halo with different masses, so they can be counted simultaneously in different mass bins. This arises not because our halos overlap physically, but due to halo and gas peculiar motions, and the finger of God redshift smearing of both halo and SBLAs.

Fig. \ref{fig:psbla sdss random} shows $\pmh$ for our three standard flux limits and our two redshifts. It is immediately clear that these SBLA samples broadly favour a specific range of halo masses, irrespective of our selection, as the peak of these distributions lie in the halo mass of $10^{11.5} \: M_{\odot}$ or higher. This tells us that at least in these two epochs (and in between them as well), SBLAs of this spectral size all broadly trace the same systems.

We estimated errors by performing a simple bootstrap analysis. Firstly, we resampled our SBLA samples 100 times. We then calculated $\pmh$ for each bootstrap realisation and estimated the uncertainty using the standard deviation of the distribution of calculated $\pmh$ values. This gives us the grey shaded areas around the cyan lines present in Fig. \ref{fig:psbla sdss random}. Overall, statistical errors are low, barely visible in the aforementioned figure, showing how tight our distribution of masses are for each SBLA sample.

We can compare this favoured halo mass to the inferred halo mass from SDSS-IV/eBOSS DR16 data in \citetalias{morrison24}, which has a range of $10^{12} - 10^{12.3} \: M_{\odot}$, shown in Fig. \ref{fig:psbla sdss random} as the vertical grey shaded area. To estimate this mass, they cross-correlated their more robust FS0 sample with the entire eBOSS DR16 $\laf$ forest sample, the amplitude of which provided an SBLA bias constraint of $b=2.34\pm0.06$. Integrating a halo mass function, one can infer from this a halo mass, which is significantly higher than the majority of the halo masses traced by the SBLAs studied in this simulation work. However, they must assume a halo mass function slope and a mass range over which to integrate. Perhaps, more importantly, the correlation function has an effective weighting, and so an effective bias. On the other hand, the mean halo mass for our samples will be dominated by the high mass end (note the logarithmic scale of our halo masses) and so these different averages both produce a higher mass than the one that can be inferred from Fig. \ref{fig:psbla sdss random}. For $z=3$ and the $\flya < 0.05$ case, we find that SDSS SBLAs trace a mean halo mass of $\approx 10^{11.78\pm0.38} \: M_{\odot}$, in broad agreement with \citetalias{morrison24} estimate, where our uncertainty represents the 1$\sigma$ deviation.

The median halo mass is 0.7 -- 0.9 dex lower (see red points in Fig. \ref{fig:hierarchy median mass}) offering a potential explanation for why bias-based estimates halo mass for  SBLAs and DLAs \citep{perez-rafols2018, perez-rafols2023} are surprisingly high with respect to what one might expect \citep{bird2014}; the bias measurements leads one to infer a mean halo mass while the median halo masses are lower by around an order of magnitude. See Sect. \ref{subsec:hierarchy} for a more extensive discussion of how SBLAs relate to halo mass.

Instead of looking at this from the point of view of the absorption systems, we can look at it from the point of view of the halos. In Eq. \ref{eq:cfrac}, we define a probe we call the `covering fraction' ($\cf$) in order to understand how SBLAs cover the projected areas of halos.
\begin{equation}
    C_{Frac} = \frac{\text{Numbers of lines-of-sight with SBLAs in a halo} }{\text{Total number of lines-of-sight in a halo} }
    \label{eq:cfrac}
\end{equation}

We recall that we remove the lines-of-sight with DLAs from the total number of lines-of-sight inside the halo. This is calculated for each individual halo, and we require the SBLAs to be associated with the halo in the redshift direction. The results can be seen in Fig. \ref{fig:cov fraction sdss random}.

The differences between redshifts and flux limits are evident upon a first viewing. As we increase the redshift and the flux limit, covering fractions get higher and higher: this is expected, as samples become larger and more inclusive. Considering both the probabilities and the covering fractions, we can state that SDSS SBLAs at lower redshifts frequently trace halos, with the downside that as we increase the flux limit, many also trace the IGM. This can be viewed as a trade-off of purity versus completeness.

It is also worthy of note that, for all samples, we can see that the covering fraction once again peaks and maintains a generally high value (in each sample) for masses $\gtrsim 10^{11.5} \: M_{\odot}$, much like in the $\pmh$ statistic we used beforehand. This once again points towards the fact that SDSS SBLAs seem to favour tracing these halo masses. However, if we vary their spectral size, would they still trace the same structures?

\section{Variable SBLA Spectral Size}\label{sec:variable size}

We have explored the relationship between SDSS SBLAs studied in the literature (\citetalias{pieri10}; \citetalias{pieri14}; \citetalias{morrison24}) and dark matter halos. As their name implies, these SBLAs are limited by SDSS resolution, and we aim to broaden our analysis to explore more SBLA samples made possible without this restriction.

In order to do this, we used the resolved spectra and rebinned them on 15 distinct scales in the range $54\kms \le v_{\text{SBLA}} \le 483\kms$ in $\approx 30\kms$ increments, to reflect 15 different SBLA spectral sizes for this study. The minimum corresponds to the typical full-width half-maximum of the spectral resolution in the blue portion of the low resolution mode of the WEAVE instrument \citep{weave_inst}, and the WEAVE-QSO survey (\citealt{weave_survey}, \citealt{weaveqso_survey}, Pieri et al. in prep.). The maximum is selected to reflect the challenge of identifying sub-DLA damping wings in real data on $\approx 6 \: \AA$ scales and to provide sufficient statistics in our synthetic spectra. We find between $10^3-10^7$ SBLAs at $z=2$, and between $10^4-10^7$ at $z=3$ (for more details, we refer readers to Table \ref{tab:sblacounts keck}). Much like the SDSS SBLAs, we also calculated the incidence rates for each spectral size and flux transmission, further explained in Appendix \ref{appendix:sbla props}.

\begin{figure}[h]
    \centering
    \includegraphics[width=0.9\linewidth]{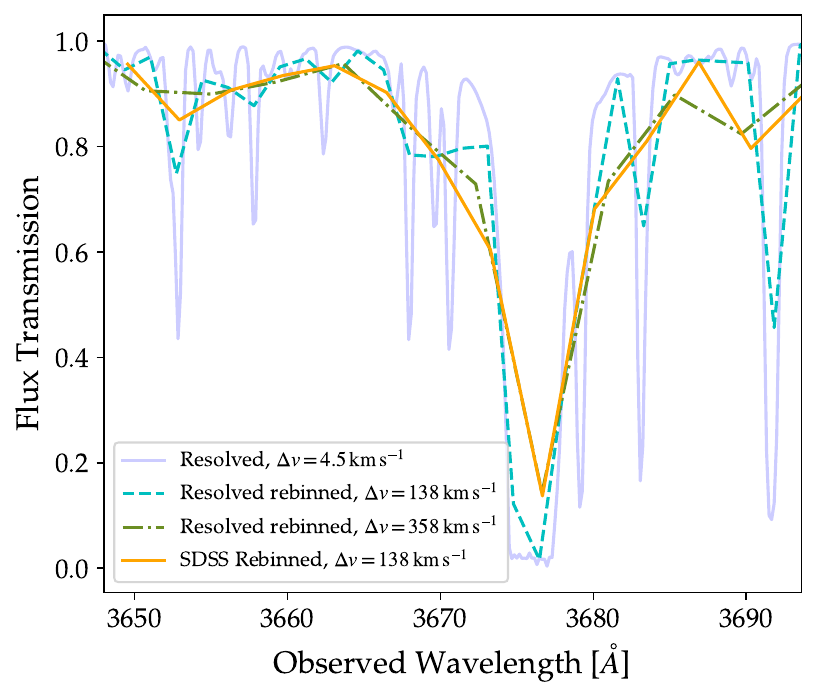}
    \caption{Comparison between a resolved spectrum rebinned to $138 \kms$ (dashed cyan line) and an SDSS spectrum rebinned to $138 \kms$ (full orange line), at $z=2$. The faded full blue line represents the resolved spectrum without any rebinning. The dot-dashed green line represents a resolved spectrum rebinned to $358 \kms$, which traces a similar structure to SDSS. There are clear differences between the rebinned resolved spectrum and the SDSS spectrum.}
    \label{fig:rebins sdss}
\end{figure}

While the strict definition for the quoted $v_{\text{SBLA}}$  (as a spectral bin size) is the same between SDSS-like and resolved spectra, the blending scale they probe is different. This is because  spectral smoothing (due to lower resolution) and binning both apply an effective smoothing to the data. The additional contribution of reduced resolution in addition to  $138 \kms$ binning leads to sensitivity to larger absorbing structures than binning alone. Fig. \ref{fig:rebins sdss} illustrates this by showing a resolved spectrum that has been rebinned to $138 \kms$ and an SDSS-like spectrum rebinned to $138 \kms$, and they look quite different; in fact, the resolved spectrum rebinned to $358 \kms$ looks similar to the SDSS spectrum. It is clear that smoothing plays an important role when it is at least of order as big as the bin size. We discuss this in more depth in Sect. \ref{subsec:smoothing}.

\subsection{Total Probability for varying SBLA Spectral Size}

\begin{figure}[h]
    \centering
    \includegraphics[width=0.85\linewidth]{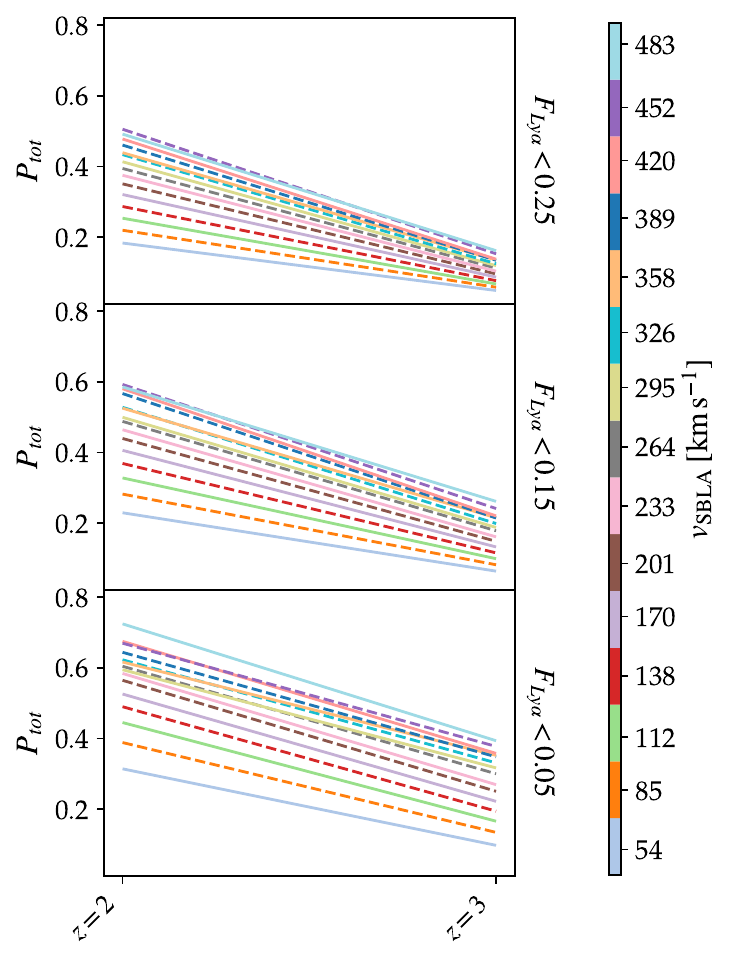}
    \caption{$\ptot$ of the variable SBLA spectral size samples, varying according to redshift. The colourbar on the side shows which line is which $\vsbla$. The full and dashed lines serve only as a visual aid.}
    \label{fig:Ptot hres nohi}
\end{figure}

Like in the previous section, we estimated the total probability of finding SBLAs in halos by using Eq. \ref{eq: psbla} for each spectral size probed. In Fig. \ref{fig:Ptot hres nohi} we can see the distribution of $\ptot$ with redshift, flux limits and SBLA spectral size. As we increase the spectral size of the SBLA sample, we find less of these objects but, contrastingly, we are more likely to find them in halos. As $\ptot$ increases, values range between $18-72\%$ at $z=2$ and $5-39\%$ at $z=3$ (see Table \ref{tab:sblacounts keck} for more details). Similarly to the SDSS SBLA samples, we also find that at $z=3$ we have roughly half the probability of finding halos than at $z=2$, although this still means that thousands of halos are being found. Furthermore, the tendency we've seen with the SDSS SBLAs of lower flux limits and lower redshift samples having higher probabilities still holds true here as well.

This is clear evidence that, besides flux limit and redshift, the spectral size of the SBLA plays a major part as well. However, as mentioned in the previous section, looking only at $\ptot$ gives us an incomplete picture of the kind of halos we trace, as these objects are more likely to trace halos of a specific mass range.

\subsection{Halo Mass Dependence for varying SBLA Spectral Size}

\begin{figure}[h]
    \centering
    \includegraphics[width=0.85\linewidth]{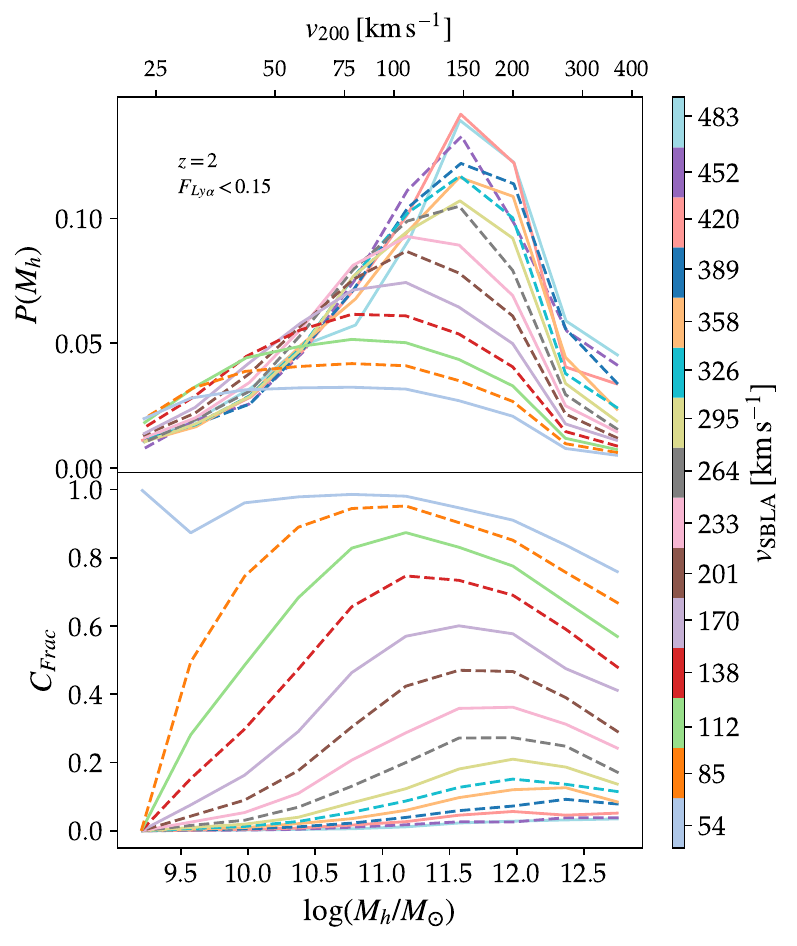}
    \caption{$\pmh$ (top panel) and $\cf$ (bottom panel), for the variable SBLA spectral size. This is an example at $z=2$ and $\flya < 0.15$. The colourbar on the side shows which line is which $\vsbla$. The full and dashed lines serve only as a visual aid.}
    \label{fig:keck psbla cfrac example}
\end{figure}

We calculated the $\pmh$ for each SBLA spectral size, in the same mass bins as mentioned for the SDSS 2-pixel case (see Table \ref{tab: halo number per mass bin}). An example of $z=2$ and $\flya < 0.15$ can be seen in the top panel of Fig. \ref{fig:keck psbla cfrac example} (for all cases, we refer readers to Appendix \ref{appendix:nohierarchy sbla}). We can see that as we increase the spectral size of the SBLA sample, the distribution changes, and the curves themselves become more peaked and favour a narrower range of masses.

It is particularly notable that, as we vary the SBLA spectral size, the peak of each distribution (or the favoured halo mass range) shifts towards higher values for each sample. This tells us that we can trace different halos by using different $\vsbla$ samples.

\begin{figure*}[h]
    \centering
    \includegraphics[width=0.87\linewidth]{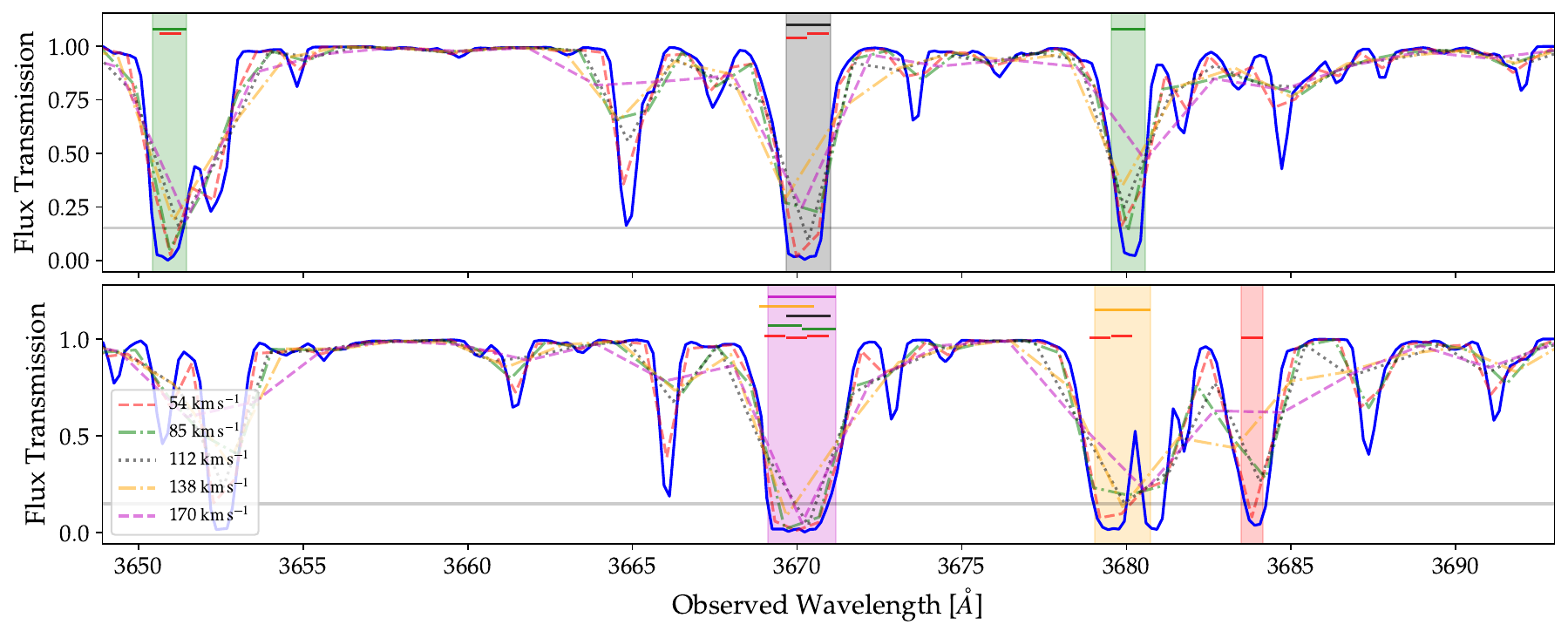}
    \caption{Example of how the hierarchy of SBLAs works. These are examples at $z=2$ and $\flya < 0.15$. The horizontal grey line represents $\flya = 0.15$. The blue full line represents the spectrum in its full resolution. The red dashed, green dot-dashed, black dotted, yellow dot-dashed and magenta dashed lines represent the spectrum rebinned to $54, 85, 112, 138$ and $170 \kms$, respectively. The red, green, black, yellow and magenta horizontal bars represent the spectral size of SBLAs with $\vsbla = 54, 85, 112, 138$ and $170 \kms$, respectively. The vertical coloured bars represent the size of the independent SBLAs.
    \textit{Top:} On the left, we can see that the SBLA with $85 \kms$ fits one SBLA with $54 \kms$ inside the same absorption feature, while on the middle the SBLA with $112 \kms$ can fit two SBLAs with $54 \kms$. \textit{Bottom:} On the right, we can see that, besides the SBLAs with $54 \kms$, only the SBLA with $138 \kms$ is detected, a product of the blending of complex absorption structure present.}
    \label{fig:hierarchy example}
\end{figure*}

But what about the halo-centric point of view? We once again calculate $\cf$, where we can see the $z=2$ and $\flya < 0.15$ case in the bottom panel of Fig. \ref{fig:keck psbla cfrac example} (for all cases, we refer readers to Appendix \ref{appendix:nohierarchy sbla}). At $z=2$, we can see that the smallest SBLAs ($< 112\kms$) cover a large range of halos masses, but as we increase $\vsbla$, we start to see that the covering fraction lowers significantly, with the larger SBLAs ($> 264 \kms$) having nearly no association with lower mass halos. This naturally follows from our previous discussion on $\pmh$ (Sect. \ref{subsec: SDSS halo mass}) where, although smaller SBLAs are numerous and trace a significant amount of halos, they also trace many parts of the IGM, lowering our probabilities but increasing the covering fraction. Meanwhile, larger SBLAs are rarer (even if still quite abundant) and exist in fewer lines-of-sight, but have a high probability of being associated with halos. Essentially, we have a purity and completeness trade-off between SBLA spectral size, covering fraction and probability, which we explore with more detail in Sect. \ref{sec:sample selection}.

Furthermore, as we increase the spectral size, the peak of the $\cf$ shifts towards higher halo masses as well, reinforcing the hypothesis that SBLA spectral size and halo mass are tightly linked.

An important ambiguity of the results set out in this section arises due the self-evident fact that extended absorption seen in wide blends (characterised by high $\vsbla$) will frequently be composed of smaller blends of strong absorption (i.e. low $\vsbla$). Small SBLAs are numerous, partly because strong absorption over a narrower velocity window is more common (while larger binning naturally draws the forest transmission to the mean). They are also numerous because several can exist within a single larger SBLA. In other words, a particular blending scale thus far (including the work of \citetalias{pieri10}, \citetalias{pieri14} and \citetalias{morrison24}) effectively defines a minimum SBLA width that includes higher $\vsbla$ absorption systems. In the next section we aim to address this ambiguity and assign SBLAs a correct and unique $\vsbla$, and in doing so improve both their veracity and utility.

\begin{figure}[h]
    \centering
    \includegraphics[width=0.85\linewidth]{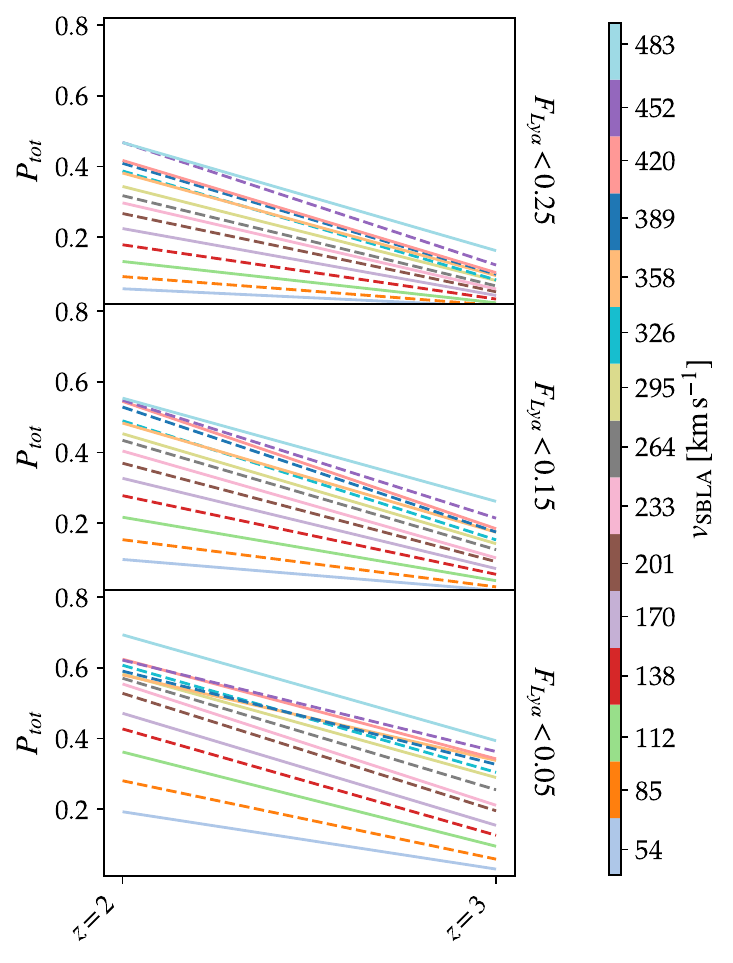}
    \caption{$\ptot$ of the variable SBLA spectral size samples, with hierarchy, varying according to redshift. The colourbar on the side shows which line is which $\vsbla$. The full and dashed lines serve only as a visual aid.}
    \label{fig:Ptot hres hi}
\end{figure}

\subsection{Hierarchical Framework}\label{subsec:hierarchy}

In the previous section, we found that SBLAs seem to associate with halos in a manner such that the halo masses are sensitive to spectral size, but this neglects the fact that extended absorbing structures are made of potentially numerous smaller ones. In many cases smaller SBLAs must be sub-divisions of larger SBLAs. In order to address this, we explore the hypothesis that, when SBLAs on different size tiers overlap, the largest SBLAs trace the dominant structure and the smaller SBLAs within them add duplication with less physical value. This is effectively a hierarchical framework since smaller SBLAs are common but larger SBLAs take priority. An added benefit of this approach is that the size of SBLA samples built this way will more closely reflect the natural extent of their absorption complex.

In Fig. \ref{fig:halo example} we illustrate the motivation for taking this approach. The resolved absorption in the top panel shows a general excess in absorption surrounding two halos (both spatially and in redshift space). In the subsequent three panels we focus only on $\flya< 0.25$, where the top panel shows the resolved resolution, while the bottom two panels show increasingly coarse spectral binnings. The smaller binning shows $\flya < 0.25$ for both halos, but the coarsest binning only shows this the more massive halo. This indicates that the less massive halo is only identified by the smaller binning and the more massive one is best-characterised by the coarser binning. In other words, the higher (lower) mass halo shows a larger (smaller) strong absorption structure, and the hierarchy serves to assign appropriate blend sizes to structures and improve the accuracy of halo mass estimation.

In effect, we treat the $\laf$ forest as an analogue for a Gaussian linear density field in Press–Schechter formalism \citep{press-schechter}. In this formalism, perturbations growth linearly until they reach a critical density contrast indicating collapse into a halo. The halo size is defined by smoothing-scale window function. Smaller perturbations begin to reach collapse before larger perturbations, and so lower mass halos begin to form earlier. This theoretical model is quasi-static in the sense that these halos form continually but do not change position. They do not merge in a physical sense, but it is possible to track how they are consumed as larger perturbations reach the collapse criterion \citep{scannapieco2001, pieri2007}. This provides a quasi-static hierarchical halo merger tree that is not physical, but that has nevertheless proven to be useful. The approach we take here is inspired by that hierarchical framework in one-dimension where a critical density contrast on a smoothing scale is exchanged for a $\laf$ flux transmission threshold on a binning scale. Larger absorption structures that reach a given absorption strength threshold consume smaller absorption structures in a hierarchical manner.

We proceed by requiring that larger SBLAs consume the smaller ones (for each redshift and flux limit), until only the independent ones remain. To do this, we go spectrum by spectrum, starting with the largest SBLAs and searching for 
SBLAs centred within their boundaries with any smaller scale, in which case they are considered consumed and are discarded. We proceed down the SBLA hierarchy, allowing surviving SBLAs on each size tier to consume smaller SBLAs centred within their boundaries as we go.
See Fig. \ref{fig:hierarchy example} for an illustration of this method. In this plot, we can see that SBLAs are not found on all spectral sizes up to a maximum: this is partly due to the bin boundaries chosen
(see upper panel of Fig. \ref{fig:hierarchy example} for an example at $3680\:$\AA, where an SBLA with $85 \kms$ is found but $54 \kms$ SBLAs are not), and also because it arises from complex absorption structures. For this second case, high transmission windows in a larger absorbing structure may represent a poor match to the intermediate scale while still providing an SBLA detection on smaller and larger scales. An example of this is seen at $3680\:$\AA\ in the lower panel of Fig. \ref{fig:hierarchy example}, where $54 \kms$ and $138 \kms$ SBLAs are found but $86 \kms$ and $122 \kms$ SBLAs are absent.

In Fig. \ref{fig:Ptot hres hi}, we recalculated $\ptot$ for every SBLA spectral size, flux limit, and redshift samples. As expected, we find that, besides the largest $\vsbla$ sample (which suffers no change), the number of detected SBLAs as well as their associated $\ptot$ has decreased amongst all samples, ranging between a $12\% - 82\%$ and $5\% - 83\%$ decrease, respectively (see Table \ref{tab:sblacounts keck} for more details). All SBLA samples still find thousands of halos.

\begin{figure*}[h]
    \centering
    \includegraphics[width=0.85\linewidth]{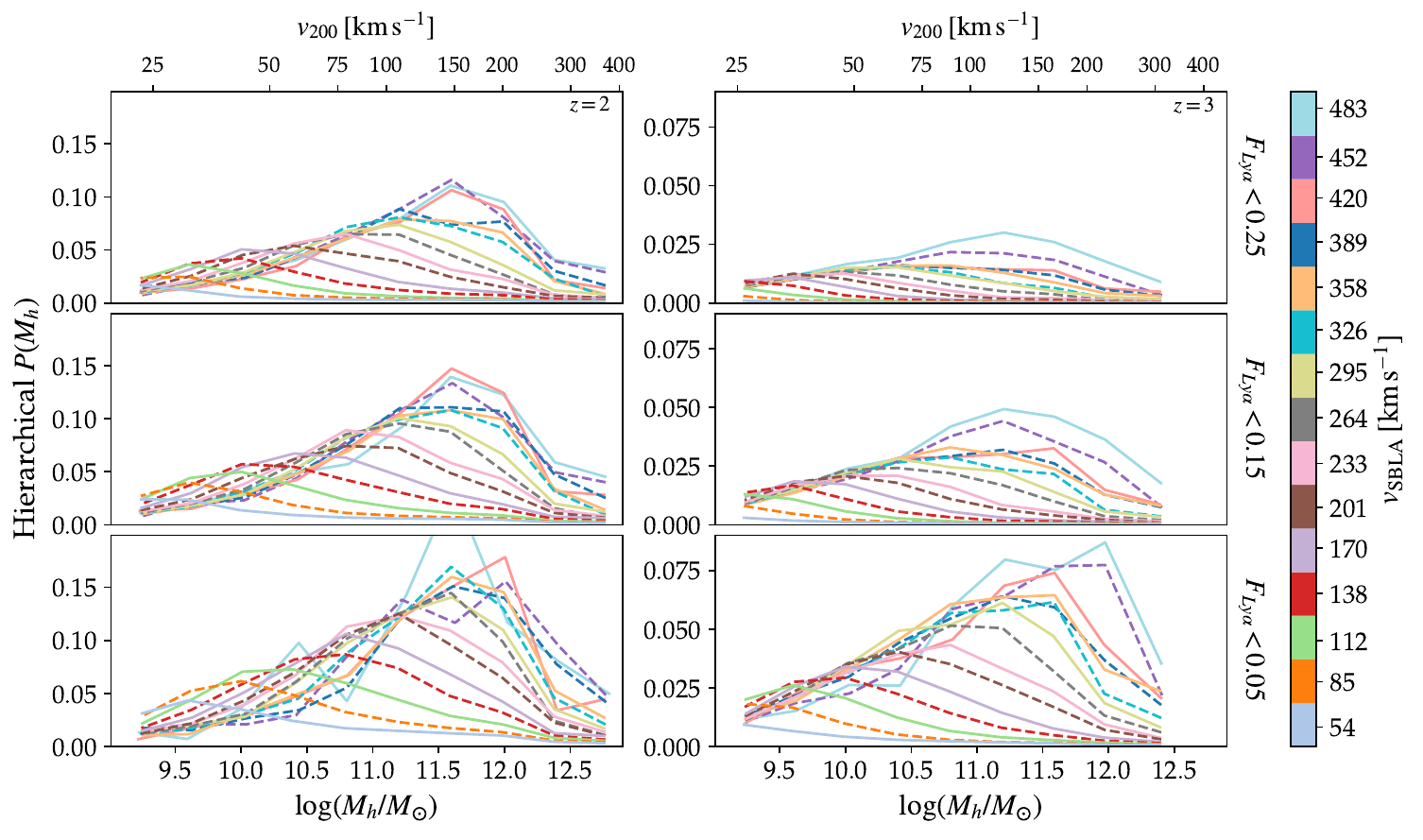}
    \caption{Hierarchical $\pmh$. $z=2$ results are on the left panels and $z=3$ results are on the right. Flux limits decrease from top to bottom. The colourbar on the side shows which line is which SBLA spectral size. The full and dashed lines serve only as a visual aid.}
    \label{fig:hierarchy psbla keck random}
\end{figure*}

In Fig. \ref{fig:hierarchy psbla keck random} we calculated $\pmh$ for this hierarchical list of objects, in order to understand the performance of the new halo mass estimation scheme. The hierarchical framework shows a clear distinction from the previous $\pmh$ estimations (see top panel of Fig. \ref{fig:keck psbla cfrac example}), where the broader distributions become narrower, and the halo mass that each $\vsbla$ sample favours is now clearer. For the smaller SBLAs ($\vsbla < 170 \kms$), we can see that they do not trace as many higher mass halos as they did before, further showing us that the hierarchical framework is effective at distinguishing between halo masses and spectral sizes. Furthermore, although larger SBLAs trace a wide range of halo masses, the peaks of their distributions have become more distinct.

\begin{figure*}[h]
    \centering
    \includegraphics[width=0.85\linewidth]{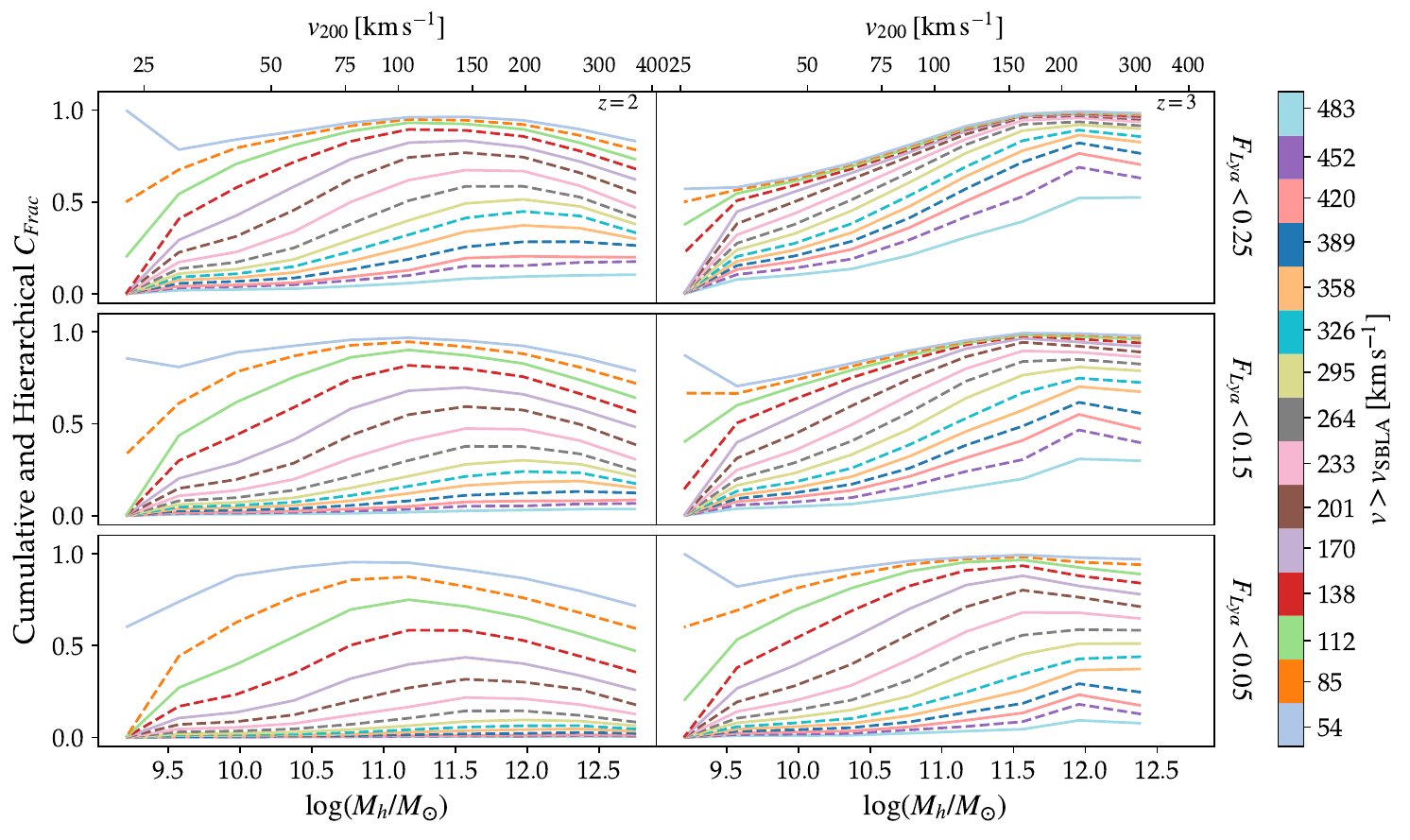}
    \caption{Covering fraction of SBLAs in the hierarchical framework, divided in mass bins. $z=2$ results are on the left panels and $z=3$ results are on the right. The colourbar on the side shows which line is which cumulative $\vsbla$ size: the full dark blue line represents $\cf$ of the $54 \kms$ SBLA and all the above, the dashed light blue represents the $\cf$ of the $85 \kms$ SBLA and all the above, and so on until we only have the $483 \kms$ SBLA spectral size. All remaining elements are the same as in Fig. \ref{fig:hierarchy psbla keck random}. The full and dashed lines serve only as a visual aid.}
    \label{fig:cov fraction keck random hierarchy}
\end{figure*}

The covering fraction of SBLAs in the hierarchical framework can be found in Fig. \ref{fig:cov fraction keck random hierarchy}. To estimate this statistic, we take a cumulative approach, where each line represents the covering fraction of a $\vsbla$ sample and all those that are larger. We chose this approach because in this framework the SBLAs on all scales must be considered as a group with respect to how effectively they identify (or fill) a population of halos. The lower limit is a free parameter because the limiting factor for a dataset will be set by the resolution of the data and its capacity to allow small SBLAs to be found. The same patterns seen in the non-hierarchical framework are present here, where the larger SBLA samples are less common and are not found in many lines-of-sight, though they are almost always in halos, while smaller SBLA samples are prevalent in most halos, but are diluted by also tracing many IGM systems. Furthermore, we also see that each hierarchy favours higher halo mass as we limit ourselves to larger and larger SBLAs,
as we saw in the hierarchical $\pmh$.

\begin{figure*}[h]
    \centering
    \includegraphics[width=0.85\linewidth]{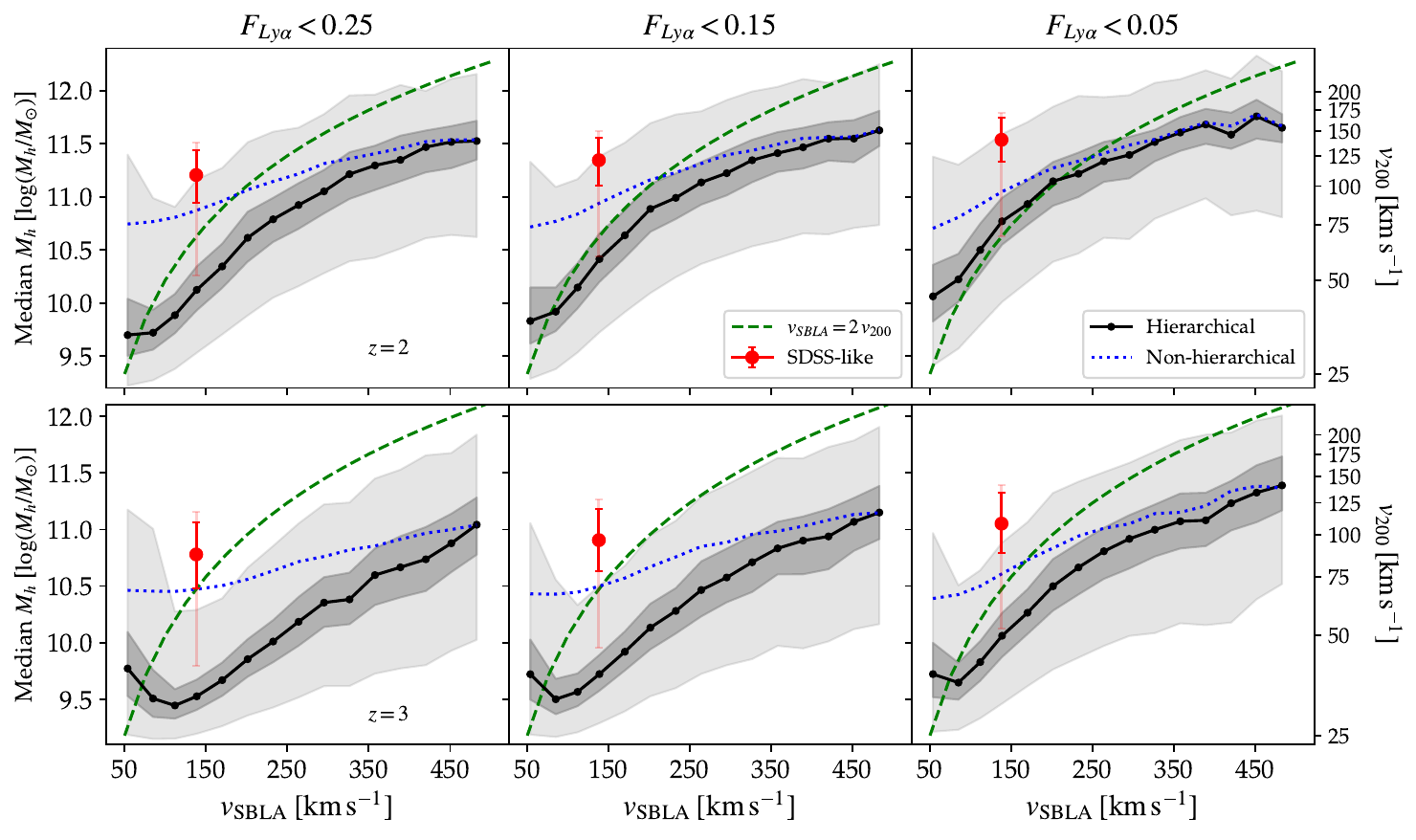}
    \caption{Median halo masses associated with SBLAs of varying spectral size. $z=2$ results are on shown in the top panels and $z=3$ on the bottom. Flux limits are reduced progressively in panels from left to right. The black line shows the median halo mass for each SBLA size in the hierarchical framework with the dark (light) grey shaded area representing the 25th and 75th percentiles (16th and 84th percentiles) spread. The dotted blue line represents the median of the non-hierarchical sample. The red points represent the SDSS SBLA samples, with the dark (light) red error bar representing the 25th and 75th percentiles (16th and 84th percentiles). The green dashed line represents the $\vsbla = 2\, \vth$ hypothesis (see Sect. \ref{subsec:velocity and sbla}).}
    \label{fig:hierarchy median mass}
\end{figure*}

The combination of halo finding probability and halo covering fraction tells us that there is a clear relationship between halo mass and SBLA spectral size. We find halos with a distinctive mass range dependent on which spectral size we consider. We note, however, that hierarchies including smaller SBLA samples are more often diluted by IGM features, notably at $z=3$. Analysing the placement and associated absorption with this suite of SBLA samples will allow us to probe different halo masses and the different properties of the CGM of their associated galaxies.

In order to understand what is the typical halo mass associated with an SBLA sample, we calculated the median mass of each $\vsbla$ sample, whose results can be seen in Fig. \ref{fig:hierarchy median mass}. It is important to note that, when calculating this, we considered only masses below $10^{13} \: M_{\odot}$ for $z=2$ and below $10^{12.6} \: M_{\odot}$ for $z=3$, given the limited statistics for larger halos. We can see that the overall trend of increasing the mass as we increase the spectral size of the absorbers is present in most flux limits for both redshifts. Relevant to note here is that the relationship between median halo mass and $\vsbla$ in the hierarchical SBLA samples trace a wider range of masses (nearly 2 dex at $\flya < 0.25$ and 1.5 dex at $\flya < 0.05$) than its non-hierarchical counterpart (about 0.9 dex at all flux limits, see blue dotted line in the aforementioned figure), further reinforcing the fact that with this framework we are better able to use $\vsbla$ to pick out different typical halo masses.

As mentioned previously, the smaller SBLAs samples are more often encroached upon by IGM systems, so one should take caution when using these samples to trace the CGM. Specifically for  $\vsbla < 90 \kms$ regime, the halo mass scatter is simultaneously large and the overall probabilities of halo finding are low. It appears that the capacity of SBLAs in finding halos breaks down for the smallest SBLA samples, particularly at $z=3$ where contamination is more common. The failure is remediated somewhat by limiting oneself to the strongest absorption possible ($\flya < 0.05$), but we discuss the limitations of these samples in Sect. \ref{subsec:velocity and sbla}.

\section{Discussion}\label{sec:discussion}

The exploration of strong-blended $\laf$ systems in \citetalias{pieri10}, \citetalias{pieri14} and \citetalias{morrison24} demonstrated that SBLAs offer an expanded set of halos in absorption that can be used to find and study the circumgalactic medium. In this work, we examined the halo finding probability, the halo covering fraction and the typical halo mass traced by SBLAs using the TNG50 simulation, using the sample by the aforementioned authors and by expanding on their work and allowing for samples with variable spectral size. We introduced the `SBLA hierarchical framework' to the community as a tool for decomposing $\laf$ forest spectra into a series of halos surrounded by IGM absorption.

In Sect. \ref{subsec:SDSS probs} we discussed the halo finding probability we derived and compared this with the observational inferences made by \citetalias{pieri14} and \citetalias{morrison24}. In order to further examine the relationship between SBLAs and halos, in this section we look at many pertinent topics left to address: uncertainties arising from our use of hydrodynamic simulations; the effects of smoothing in the spectra; the relationship between SBLA spectral size, halo mass, and halo circular velocity; how to choose a sample to study the CGM or the IGM; and discuss our definition of the boundary of the CGM.

\subsection{Hydrodynamic Simulation Limitations}\label{sec:hydro lims}

Although the TNG50 simulation possesses exquisite resolution and is able to provide us with many details about the Universe, there are still some caveats we must address with our results. The first one is that the TNG50 simulation does not resolve the parsec-scale clumping associated with SBLAs as inferred in \citetalias{pieri10}, \citetalias{pieri14} and \citetalias{morrison24}. The second is that we have rescaled the mean flux of the spectra at each redshift in order to ensure that they follow the measurements of \cite{fauchergiguere08}.

These two factors imply that the synthetic spectra that we used to select SBLAs may not be perfectly representative of observations. Any such small-scale structure would impact metal absorption properties, and may also have an impact on associated $\laf$ absorption. Meanwhile a global mean flux consistent with observations does not guarantee obtaining the correct representation for the $\approx 1\%$ of the forest associated with CGM absorbers. Both these subtleties may have an impact on our results.

As discussed in Sect. \ref{subsec:SDSS probs}, we have noticed a slight discrepancy between the $\ptot$ of our SDSS SBLAs in the $z=3$ box with the \citetalias{pieri14} and \citetalias{morrison24} measurements. One way to understand if this is a by-product of any simulation limitations is by comparing incidence rates, $dn/dz$, obtained from the simulation with the values from observations (see Appendix \ref{appendix:sbla props} for more details). We find that this discrepancy holds at $z=3$, where incidence rates in TNG50 are twice or even three times higher than what is found in rebinned KODIAQ and SQUAD spectra (so that they resemble SDSS). These results may perhaps indicate tensions with the TNG50 simulation results, its physical model, our treatment of the UV Background and neutral hydrogen modelling, and/or our synthetic spectra post-processing. Even with this in mind, this does not detract from the fact that we are broadly able to replicate results from \citetalias{morrison24}, namely the halo mass estimate from the SDSS-like spectra, even using the $z=3$ box.

\subsection{Smoothing and Noise Effects}\label{subsec:smoothing}

We have thus far treated the SBLA spectral size derived from the resolved spectra as a variable that can be applied to any case, i.e. any resolution or instrument. However, when we compare Fig. \ref{fig:psbla sdss random} with Fig. \ref{fig:keck psbla cfrac example}, we can see that this is not the case: the SDSS SBLA sample, although technically possessing a $\vsbla = 138\kms$, does not behave in the same way as the resolved SBLA sample of the same spectral size. The reason for this traces back to the fact that the SDSS spectra are smoothed by instrumental broadening on a scale that exceeds the wavelength solution, with the full width at half-maximum (FWHM) varying between $138-179 \kms$ \citepalias{pieri14}, changing the effective spectral size of the SBLAs that we find.

In order to compare how the smoothing affects the estimates from SBLAs, we can look at the median halo mass estimates from the resolved SBLA samples and the SDSS SBLA samples, visible in the black lines and red dots in Fig. \ref{fig:hierarchy median mass}, respectively. The differences between the $\vsbla$ in SDSS and the resolved spectra become clear here as, for the same size, two clearly distinct halo masses are preferred, albeit with significant overlap in the spread. Effectively, this means that the SDSS-like SBLAs are comparable with the resolved SBLAs that have $\vsbla \gtrsim 320 \kms$. As such, a connection between these two cases should then be made with these spectral sizes, and not with a direct one-to-one relation.

This is not unexpected: in fact, the combination of smoothing and binning implies larger absorbing structures than binning alone. We will further explore the role of smoothing and binning in SBLA sample building and subsequent halo finding in Hu et al. (in prep.).

Another important effect that requires further comment is the presence of noise. Most data present in large spectroscopic surveys (such as SDSS or DESI) have low-to-moderate signal-to-noise ratios. This will clearly have an important impact on the detection and measurement of SBLAs. \citetalias{pieri14} studied mock data with added noise. As we noted in Sect. \ref{sec:SDSS-like}, they found that the $\flya < 0.05$ SBLA samples with noise included, for $\text{S/N/\AA} = 3$, resembles the  true $\flya < 0.15$ samples (see their Fig. 3). \citetalias{morrison24} also argued that, by training on mocks, one can generate fixed purity samples to some true maximum flux by varying the flux limit based on the local signal-to-noise. These studies were focussed on SDSS data binned to $138\kms$, and noise effects on higher resolution data are likely to have some quantitative differences, but the general qualitative picture remains that one can require stronger absorption to boost purity in light of observing noise. Once again, a more extensive exploration of this will be presented in Hu et al. (in prep.).

\subsection{A Toy Model for SBLA Spectral Size, Halo Mass and Halo Circular Velocity}
\label{subsec:velocity and sbla}

In addition to the inferred mean halo mass derived from a forest cross-correlation (as discussed in Sect. \ref{subsec: SDSS halo mass}), \citetalias{morrison24} put forward a hypothesis as to why SBLAs are associated with halos. The picture proposed is one where halos are filled with optically thin gas clouds (or some smoothly varying gas distribution equivalent) that trace the dark matter dynamics. 

Specifically, one must imagine a sphere of radius equal to the virial radius with gas circling randomly with velocities of up to the virial velocity such that any line of sight may encounter gas moving with velocity between $+\vth$ and $-\vth$. Indeed there is some support for this picture from the distribution of line-of-sight velocities in halos. For example, in Fig. 3 of \citealt{weng24}, we can see that even in evolved systems we have components that possess the same velocity as the halo circular velocity -- equivalent to our $\vth$. 
\citetalias{morrison24} goes a step further and suggests that, at $z>2$, strong $\laf$ absorption might fill this kinematic window such that (in combination with broadening) it fills a spectral range equivalent to $\vsbla = 2\,\vth$. If this picture were correct it offers the tantalising prospect of estimating halo masses directly from the absorbing structure.
This hypothesis is shown as the green-dashed line in Fig. \ref{fig:hierarchy median mass} allowing us to examine its validity and the insights it may provide.

We require SBLAs to be associated with halos if they are inside the viral radius (see Eq. \ref{eq:rvir}) and within $2\,\vth$ in velocity space (see Eq. \ref{eq:deltav}). This requirement in itself does not provide that $\vsbla$ will equal twice $\vth$; the above hypothesis is suggests that the kinematic window should be filled with absorbing gas but the specific spectral size will depend on how the size is measured and the details of how gas dynamics maps to a $\laf$ absorption.
Note also that we use hard limits on the decision of halo in/out and SBLA boundaries here, but halos are not hard edged (see Sect. \ref{subsec:CGM region}) and in SDSS the Gaussian instrumental broadening clearly plays a role (as explained in Sect. \ref{subsec:smoothing}).

Turning back again to Fig \ref{fig:hierarchy median mass}, we compare the median halo mass estimate, in the black line, with the hypothesised value of $\vsbla = 2\,\vth$, in the green dashed line. At $z=2$, we can see that, as we lower the flux limit, our estimates increasingly align with this hypothesis, with failure to conform to this expectation coming from the smallest SBLAs ($\vsbla \leqslant 85\kms$). In any case, most $\vsbla$ samples reside below this line, implying that this hypothesis is a good indicator of the upper limit of the halo mass that SBLAs trace. For $z=3$, this is also true, with the caveat that once again the smallest SBLAs do not conform to the expectation of the hypothesis. 

The SDSS SBLA samples at both redshifts seem to match the $\vth$ directly, with no need for the adjustment to the $2\, \vth$, no doubt due to the smoothing kernel.

\subsection{CGM and IGM Sample Selection} \label{sec:sample selection}

With the hierarchical framework, we found that there is a striking relationship between SBLA spectral size and typical halo mass traced, something that is not as evident in the non-hierarchical case. With this link established, the next step is to understand which sample of SBLAs we should use to study the CGM regions associated with these halos.

Of course one would like to build a CGM sample with an acceptable level of both purity and completeness, and the details of how one does this depends on both what is possible and the specific requirements of the study. In the previous section, we found that there is some small SBLA limit for which the probability that an SBLA finds a halo is rather low (see Fig. \ref{fig:hierarchy psbla keck random}). Hence, it is natural to build a hierarchical SBLA sample down to some minimum $\vsbla$ to study a range of halo masses down to some chosen failure threshold. We therefore study purity and completeness as top-down cumulative statistics. Specifically the cumulative purity is the probability of finding any SBLA above a certain size, and the probability in question for any $\vsbla$ is the hierarchical $\ptot$ (see Fig. \ref{fig:Ptot hres hi}).

Regarding completeness, we assess all lines-of-sight in halos with SBLAs of a certain size and above (a cumulative SBLA size, much like the purity) and divided them with all lines-of-sight in halos. For these estimates, we once again restricted the halo masses to the range defined in Sect. \ref{subsec: SDSS halo mass}, where at $z=2$ we consider $10^{9} \: M_{\odot} < M_h < 10^{13} \: M_{\odot}$ and at $z=3$ we consider $10^{9} \: M_{\odot} < M_h < 10^{12.6} \: M_{\odot}$. This is similar to the covering fraction, but instead of having a mean per halo, we consider it a global statistic describing all lines-of-sight.

\begin{figure}
    \centering
    \includegraphics[width=0.85\linewidth]{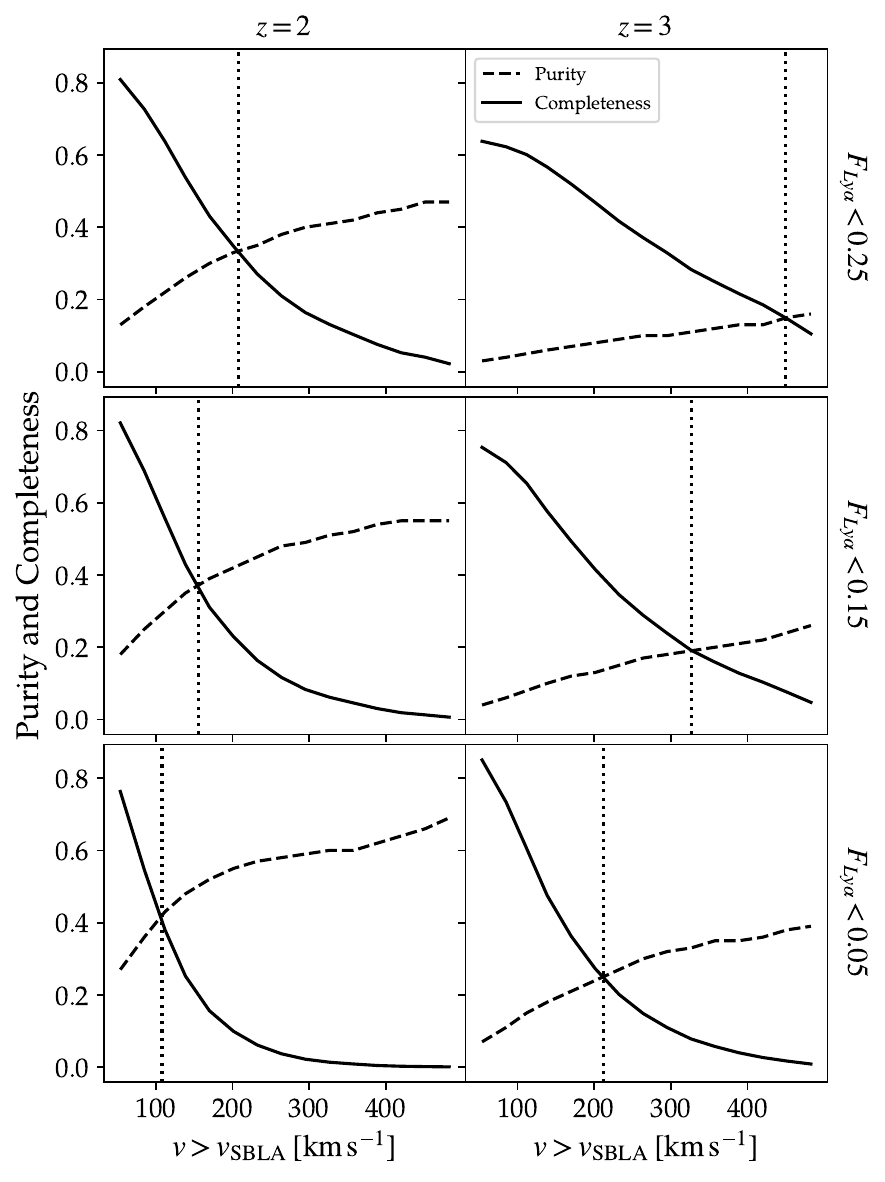}
    \caption{Purity (or the cumulative $\ptot$ for the hierarchical SBLAs, dashed line) and completeness (full line). The dotted vertical line represents the point where purity and completeness intersect. On the left hand panels, we have $z=2$, and on the right we have $z=3$, with the flux transmission limits decreasing from top to bottom.}
    \label{fig:purity completeness}
\end{figure}

The cumulative purity and completeness  are shown in Fig. \ref{fig:purity completeness}. In order to construct a sample of CGM systems to study, a good rule of thumb is to balance purity and completeness by looking for the point of equality. This provides us with a full SBLA sample based on a hierarchy with some minimum $\vsbla$ for any given flux limit. The preferred flux limit to select is $\flya < 0.05$ given that this provides the highest purity and completeness equality. Applying this to data in surveys such as the Dark Energy Spectroscopic Instrument (DESI, \citealt{desi_inst,desi_edr}) or the WHT Enhanced Area Velocity Explorer (WEAVE \citealt{weave_inst,weave_survey}) suveys
is in our future work, as we believe testing this in observations is of crucial importance. However, these low flux transmissions require high signal-to-noise data (in light of SBLA purity analysis of \citetalias{morrison24}). Probing halos of mass of $\approx 10^{10} M_{\odot}$ at $z=2$ therefore require  both high resolution spectra (to resolve  $\vsbla = 100 \kms$) and high signal-to-noise (to reliably recover true transmission of 5\%).

One may wish to prioritise a complete sample of CGM systems with less priority on purity in order to trace baryon acoustic oscillations with the largest available sample of halos, as was explored in \citet{perez-rafols2023}. On the other hand, one may wish to study a pure sample of IGM by excluding a more complete sample of CGM systems. For example, \citet{morrison2021} used metal stacks to study the UV background spatial variation, with the inclusion and exclusion of SBLAs. For such samples, we can choose to trade off some of the purity in the selection with a boost in completeness, by increasing the flux limits and maintaining the $\vsbla$ for the CGM sample. Higher flux limits provide a larger number of SBLAs (e.g. Table \ref{tab:sblacounts sdss in}), and allow to take into account the effects of noise (as explained when we compare our purity with \citetalias{morrison24}).

In effect, to construct a sample of CGM systems to study, at $z=2$ we suggest an SBLA selection with $\flya < 0.05$ and $\vsbla \geqslant 112 \kms$, which translates into a purity of $45\%$ and completeness of $40\%$; while at $z=3$ we suggest $\flya < 0.05$ and $\vsbla \geqslant 201 \kms$, which gives us a purity and completeness of about $25\%$. For IGM studies, or completeness prioritised CGM samples, at $z=2$ we suggest an SBLA sample with $\flya < 0.25$ and $\vsbla \geqslant 112 \kms$, which translates into a purity of $25\%$ and completeness of $65\%$. At $z=3$, suggestions for a large SBLA sample becomes more complicated, as at $\flya < 0.25$ we have low purity and completeness at nearly all $\vsbla$. With $\flya < 0.15$ we have higher purity and completeness, so we suggest this flux limit as well as $\vsbla \geqslant 201\kms$, with a purity of $18\%$ and completeness of $43\%$. 

\subsection{Boundaries of the CGM Region}\label{subsec:CGM region}

When we define the CGM in this work, we are assuming that the virial radius -- which we define as the region of space where the density is 200 times the mean density of the Universe \citep{Gunn72} -- encompasses all (or most) of the particles that are tied to the central galaxy in the halo. However, this has been shown to not always be the case (e.g. \citealt{Diemer17,garcia23}).

If we consider one of these halo boundaries to replace the virial radius, we will expand the region of physical space which will, in turn, affect our statistics. The $\ptot$ and $\pmh$ statistics (hierarchical or not) will increase, as more SBLAs will be included inside the CGM regions of galaxies. When it comes to $\cf$, the impact is more subtle, as it depends on the gas properties of the increased radius for each halo.

This is only considering extensions to the CGM definition in physical space, but the truth is that velocity space would be affected as well: $\vth$ could be an insufficient measurement to describe where SBLAs reside spectroscopically in the halos. However, there does not seem to exist a relationship between increasing the radius of a halo and increasing/decreasing the spectral velocity of the components present in it (see Fig. 9 of \citealt{weng24}), so while changing the radius of the halo is expected to have an impact in the physical space, in velocity space our results should remain unchanged. 

Finally, we note that it is standard practise in observational studies of the CGM to define it as a fixed range of velocity and impact parameter (e.g. 300 pkpc and $\pm 300 \kms$ in \citealt{rudie2012} and \citetalias{pieri14}) in the absence of a known galaxy halo mass. These CGM regions are generally larger than the ones studied here for all but the largest halos we study (approximately $10^{12} \: M_{\odot}$ in TNG50). We will study larger zones in Hu et al. (in prep.), but the broad picture is one of higher purity and lower completeness.

\section{Conclusions}\label{sec:conclusion}

Strong, Blended $\laf$ (SBLA) absorbers have already been shown to have the potential to detect many halos in the $\laf$ forest \citepalias{pieri10,pieri14,morrison24}. With this work, we study this observational phenomenon and explore which halos are traced by SBLAs. We do this by exploiting absorption spectra generated from the TNG50 cosmological hydrodynamical simulation. We possess 4 million parallel lines-of-sight, randomly spread throughout the box, with minimum wavelength of $3645 \: \AA$ ($4870 \: \AA$) and a maximum wavelength of $3695 \: \AA$ ($4930 \: \AA$), at $z=2$ ($z=3$). We study have these spectra in two resolutions: SDSS-like ($\Delta v = 69 \kms$) and `resolved' ($4.5 \kms$).

We find that:

\begin{itemize}
    \item In SDSS resolution, SBLAs with the lower $\laf$ flux transmission ($\flya < 0.05$) have a $67\%$ ($23\%$) probability of finding halos at $z=2$ ($z=3$). A 5-15\% drop in probability occurs per 10\% increase in transmission to $\flya < 0.25$;

    \item The mean halo mass of these SDSS SBLAs is $10^{11.78} \: M_{\odot}$. This is broadly consistent with the value derived in \citetalias{morrison24} based on clustering bias ($10^{12} \; \text{to} \; 10^{12.3} \: M_{\odot}$) within their modelling uncertainty;

    \item SBLAs are more numerous and cover more halo lines-of-sight at higher redshifts and higher flux transmissions;
    
    \item In resolved spectra we are free to select the SBLA spectral size and do so between $54 \kms \leqslant \vsbla \leqslant 483 \kms$. Upon doing so we find that the probability of finding halos increases with SBLA size and reduced flux transmission;

    \item We apply a hierarchical framework, allowing larger SBLAs to consume the smaller ones (see illustration in Fig. \ref{fig:hierarchy example}). This allows SBLAs to reach their natural spectral size and avoids duplication. With this refinement the halo mass distribution for a given SBLA spectral size becomes narrower with respect to the non-hierarchical case;

    \item In the hierarchical framework, we trace median halo masses from $M_h \approx 10^{9.5} M_{\odot}$ (for $100 \kms$ SBLAs) to $M_h \approx 10^{11.5} M_{\odot}$ (for $450 \kms$ SBLAs);

    \item In order to study the CGM, we recommend selecting SBLAs with the lower flux limit ($\flya < 0.05$) and with $\vsbla \geqslant 112\kms$ ($\vsbla \geqslant 201\kms$), with purity and completeness of $45\%$ and $40\%$ ($25\%$ and $25\%$), respectively, for $z=2$, ($z=3$). For IGM studies, we suggest selecting SBLAs with $\flya < 0.25$ ($\flya < 0.15$) and $\vsbla \geqslant 112\kms$ ($\vsbla \geqslant 201\kms$), with purity and completeness of $25\%$ and $65\%$ ($18\%$ and $43\%$), respectively, for $z=2$ ($z=3$). For these examples, we recommend caution, as they have been optimised for noiseless and high resolution data.
\end{itemize}

In the future, we plan on exploring the effects of smoothing as well as the combination of smoothing and rebinning in the SBLA selection. We will also apply this hierarchical framework to both high resolution data, such as KODIAQ \citep{kodiaq}, as well as lower resolution data, such as DESI \citep{desi_inst,desi_edr} or 
WEAVE-QSO \citep{weaveqso_survey}, as part of WEAVE \citep{weave_inst,weave_survey} to understand if the results from this work hold true. In the next paper of this series, we will explore the effects of smoothing and noise in the simulated spectra, in order to understand how resolution and signal-to-noise of data impacts SBLA selection across all scales.

\begin{acknowledgements}

This work was supported by the French National Research Agency (ANR) under contract ANR-22-CE31-0026 and by Programme National Cosmology et Galaxies (PNCG) of CNRS/INSU with INP and IN2P3, co-funded by CEA and CNES.

D.N. acknowledges funding from the Deutsche Forschungsgemeinschaft (DFG) through an Emmy Noether Research Group (grant number NE 2441/1-1), and under Germany's Excellence Strategy EXC 2181/1 - 390900948 (the Heidelberg STRUCTURES Excellence Cluster).

M.R.H.B. acknowledges funding from grant CNS2022-135878 funded by MICIU/AEI/10.13039/501100011033 and by the European Union NextGenerationEU/PRTR, and from the project I+D+i PID2024-156844NA-C22, also funded by MICIU/AEI/10.13039/501100011033.

This work made use of Astropy:\footnote{http://www.astropy.org} a community-developed core Python package and an ecosystem of tools and resources for astronomy \citep{astropy:2013, astropy:2018, astropy:2022}.

\end{acknowledgements}

\bibliographystyle{aa}
\bibliography{bibliography.bib}

\begin{appendix}

\section{SBLA Counts, $\ptot$ and Incidence Rates}\label{appendix:sbla props}

In this section, we show the results from the matching of the SBLA positions within the halos, as described in Sect. \ref{sec:sbla finding}. We also calculated the incidence rate of SBLAs.

From observations, we have incidence rates estimated from 130 SQUAD \citep{squad} QSO spectra. At $z=2$, we have a redshift path length of 0.6, while at $z=3$ we have 1.4. A cut in signal-to-noise (S/N) in each pixel was applied, where we enforce $\text{S/N}$ per pixel $ > 10$, and also rebinned and smoothed the spectra to match the SDSS 2-pixel rebin wavelength solution, and we ended with 44 QSOs. Then, we  divided the data in two redshift bins: one between $1.8 < z < 2.2$ and another between $2.8 < z < 3.2$. We also have access to 150 KODIAQ \citep{kodiaq} QSO spectra. The same cut in S/N was applied, as well as the same rebin and smoothing so the wavelength solution is the same as SDSS 2-pixel rebin, leaving us with 73 QSO spectra. The data was then divided in two redshift bins: $2.15 < z < 2.42$ and $2.8 < z < 3.5$. The incidence rates of both SQUAD and KODIAQ spectra can be found in Table \ref{tab:squad dndz}. All DLAs and sub-DLAs were masked.

For the simulations, we define the incidence rate as:
\begin{equation}
    \frac{dn}{dz} = \frac{\text{Number of SBLAs detected}}{\text{Wavelength Interval} /\lambda_{\laf} \cdot \text{Number of lines-of-sight}}
\end{equation}
where the wavelength interval is either $50 \: \AA$ or $60 \: \AA$ for $z=2$ and $z=3$, respectively, the number of lines-of-sight is 4 million, minus the DLA sightlines (see Sect. \ref{sec:hydro spectra}), and $\lambda_{\laf}$ is the wavelength of the $\laf$ line, $1\,215.67 \:$\AA. The results of this can be seen in Table \ref{tab:sblacounts sdss in} for SDSS SBLAs and Table \ref{tab:sblacounts keck} for variable SBLA spectral sizes, for all redshifts, alongside all the probabilities and SBLA counts mentioned in the main text.

\begin{table}[h]
\caption{Incidence rates estimated from SQUAD and KODIAQ spectra.}
\centering
\label{tab:squad dndz}
\setlength{\tabcolsep}{8pt}
\renewcommand{\arraystretch}{1.2}
\begin{tabular}{c|c|c}
 SQUAD               & $1.8 < z < 2.2$ & $2.8 < z < 3.2$ \\
\multicolumn{1}{c|}{$\flya$} & $dn/dz$           & $dn/dz$         \\ \hline
$[0.15, 0.25)$ & $3.4 \pm 0.47$ & $10.9 \pm 1.1$ \\
$[0.05, 0.15)$ & $2.7 \pm 0.43$  & $11.3 \pm 1.1$  \\
$< 0.05$ & $1.1 \pm 0.26$  & $4.7 \pm 0.71$   \\ \hline \hline        

KODIAQ                & $2.15 < z < 2.42$ & $2.8 < z < 3.5$ \\
\multicolumn{1}{c|}{$\flya$} & $dn/dz$ & $dn/dz$  \\ \hline
$[0.15, 0.25)$ & $8.41 \pm 1.7$ & $14.2 \pm 2.3$ \\
$[0.05, 0.15)$ & $4.4 \pm 1.7$  & $7.0 \pm 1.8$  \\
$< 0.05$ & $2.0 \pm 1.2$  & $1.6 \pm 0.9$   \\ \hline         
\end{tabular}
\tablefoot{The spectra were smoothed and rebinned to be identical to the SDSS 2-pixel wavelength solution}
\end{table}

Comparing the results from Tables \ref{tab:squad dndz} and \ref{tab:sblacounts sdss in}, we can see that there is a clear difference between them. At $z=2$, we find that the incidence rate of SBLAs is typically lower than what is expected from KODIAQ, but within the limits of the SQUAD estimates, showing that at this redshift we have a degree of confidence in the simulations. At $z=3$, though, the story is quite different: we find that the $dn/dz$ from simulations is generally twice or three times as higher than what is observed in data.

This is a potential flag that there might be some issues in this generation of the spectra in the TNG50 simulation, as, for example, there could be difficulties in handling the transition between linear to non-linear structure formation at higher redshifts. On the other hand, although the SQUAD spectra have high signal-to-noise, there are still some noise effects -- strong absorption could be slightly over/underestimated and SBLAs could be put in the `wrong' $\flya$ bin, boosting the $dn/dz$ estimation of one bin and reducing the other.

\begin{table*}[h]
\caption{Total number of SBLAs, $n_{\text{SBLA}}$, the respective probability of finding them in halos, $\ptot$, for each redshift in the TNG50 simulation box.}
\centering
\label{tab:sblacounts keck}
\setlength{\tabcolsep}{8pt}
\renewcommand{\arraystretch}{1.2}
\begin{tabular}{cc|ccccc|ccccc}
$\vsbla$ & & \multicolumn{5}{|c|}{$z=2$} &  \multicolumn{5}{|c}{$z=3$} \\ 

[$\text{km} \: \text{s}^{-1}$] & $\flya$ & $n_{\text{SBLA}}$ & $\ptot$ & $n_{h, \text{SBLA}}$ & $P_{h,tot}$ & $dn/dz$ & $n_{\text{SBLA}}$ & $\ptot$ & $n_{h, \text{SBLA}}$ & $P_{h,tot}$ & $dn/dz$ \\ \hline
\multirow{3}{*}{54}& $< 0.25$ & 13$\,$268$\,$521 & 0.18 & 3$\,$310$\,$301 & 0.05 & 20.54 & 45$\,$253$\,$628 & 0.05 & 8$\,$729$\,$618 & 0.0 & 46.6 \\
& $< 0.15$ & 9$\,$501$\,$758 & 0.23 & 2$\,$951$\,$002 & 0.1 & 18.31 & 33$\,$175$\,$144 & 0.06 & 8$\,$867$\,$342 & 0.01 & 47.34 \\
& $< 0.05$ & 5$\,$498$\,$951 & 0.31 & 2$\,$214$\,$722 & 0.19 & 13.74 & 19$\,$989$\,$161 & 0.1 & 7$\,$360$\,$312 & 0.03 & 39.29 \\ \hline

\multirow{3}{*}{85}& $< 0.25$ & 6$\,$597$\,$498 & 0.22 & 1$\,$650$\,$782 & 0.09 & 10.24 & 2$\,$3867$\,$995 & 0.06 & 4$\,$301$\,$141 & 0.01 & 22.96 \\
& $< 0.15$ & 4$\,$307$\,$885 & 0.28 & 1$\,$324$\,$103 & 0.15 & 8.22 & 15$\,$866$\,$509 & 0.08 & 3$\,$842$\,$303 & 0.02 & 20.51 \\
& $< 0.05$ & 2$\,$163$\,$281 & 0.39 & 838$\,$390 & 0.28 & 5.2 & 8$\,$283$\,$265 & 0.13 & 2$\,$727$\,$084 & 0.06 & 14.56 \\ \hline

\multirow{3}{*}{112}& $< 0.25$ & 4$\,$015$\,$923 & 0.25 & 1$\,$098$\,$122 & 0.13 & 6.81 & 15$\,$464$\,$301 & 0.07 & 2$\,$912$\,$033 & 0.01 & 15.55 \\
& $< 0.15$ & 2$\,$430$\,$704 & 0.33 & 820$\,$123 & 0.22 & 5.09 & 9$\,$520$\,$623 & 0.1 & 2$\,$437$\,$582 & 0.04 & 13.01 \\
& $< 0.05$ & 1$\,$087$\,$094 & 0.45 & 464$\,$148 & 0.36 & 2.88 & 4$\,$438$\,$240 & 0.17 & 1$\,$537$\,$354 & 0.09 & 8.21 \\ \hline

\multirow{3}{*}{139}& $< 0.25$ & 2$\,$598$\,$739 & 0.29 & 806$\,$626 & 0.18 & 5.01 & 1$\,$0671$\,$222 & 0.08 & 2$\,$163$\,$993 & 0.02 & 11.55 \\
& $< 0.15$ & 1$\,$459$\,$972 & 0.37 & 551$\,$799 & 0.28 & 3.42 & 6$\,$113$\,$412 & 0.12 & 1$\,$665$\,$131 & 0.05 & 8.89 \\
& $< 0.05$ & 582$\,$547 & 0.49 & 276$\,$058 & 0.43 & 1.71 & 2$\,$573$\,$833 & 0.19 & 950$\,$712 & 0.13 & 5.08 \\ \hline

\multirow{3}{*}{170}& $< 0.25$ & 1$\,$634$\,$429 & 0.32 & 477$\,$874 & 0.22 & 2.97 & 7$\,$371$\,$416 & 0.09 & 1$\,$592$\,$813 & 0.03 & 8.5 \\
& $< 0.15$ & 853$\,$553 & 0.41 & 306$\,$296 & 0.33 & 1.9 & 3$\,$917$\,$196 & 0.13 & 1$\,$150$\,$919 & 0.07 & 6.14 \\
& $< 0.05$ & 303$\,$582 & 0.53 & 140$\,$700 & 0.47 & 0.87 & 1$\,$468$\,$194 & 0.22 & 585$\,$543 & 0.15 & 3.13 \\ \hline

\multirow{3}{*}{201}& $< 0.25$ & 1$\,$077$\,$730 & 0.35 & 362$\,$331 & 0.27 & 2.25 & 5$\,$334$\,$243 & 0.1 & 1$\,$220$\,$715 & 0.04 & 6.52 \\
& $< 0.15$ & 521$\,$254 & 0.44 & 214$\,$384 & 0.37 & 1.33 & 2$\,$636$\,$842 & 0.15 & 802$\,$922 & 0.09 & 4.29 \\
& $< 0.05$ & 162$\,$109 & 0.56 & 83$\,$775 & 0.53 & 0.52 & 886$\,$698 & 0.25 & 364$\,$054 & 0.2 & 1.94 \\ \hline

\multirow{3}{*}{233}& $< 0.25$ & 733$\,$287 & 0.37 & 244$\,$017 & 0.3 & 1.51 & 3$\,$929$\,$248 & 0.1 & 896$\,$787 & 0.05 & 4.79 \\
& $< 0.15$ & 331$\,$270 & 0.46 & 134$\,$240 & 0.4 & 0.83 & 1$\,$815$\,$055 & 0.16 & 558$\,$370 & 0.1 & 2.98 \\
& $< 0.05$ & 93$\,$507 & 0.58 & 49$\,$143 & 0.55 & 0.3 & 553$\,$897 & 0.27 & 237$\,$159 & 0.21 & 1.27 \\ \hline

\multirow{3}{*}{264}& $< 0.25$ & 505$\,$709 & 0.39 & 170$\,$725 & 0.32 & 1.06 & 2$\,$985$\,$871 & 0.11 & 702$\,$000 & 0.06 & 3.75 \\
& $< 0.15$ & 212$\,$398 & 0.49 & 88$\,$452 & 0.43 & 0.55 & 1$\,$270$\,$738 & 0.18 & 396$\,$560 & 0.12 & 2.12 \\
& $< 0.05$ & 52$\,$328 & 0.6 & 29$\,$604 & 0.57 & 0.18 & 341$\,$367 & 0.3 & 145$\,$028 & 0.25 & 0.77 \\ \hline

\multirow{3}{*}{295}& $< 0.25$ & 349$\,$131 & 0.41 & 111$\,$280 & 0.34 & 0.69 & 2$\,$324$\,$111 & 0.12 & 611$\,$884 & 0.08 & 3.27 \\
& $< 0.15$ & 136$\,$070 & 0.5 & 53$\,$649 & 0.45 & 0.33 & 940$\,$272 & 0.19 & 326$\,$311 & 0.14 & 1.74 \\
& $< 0.05$ & 29$\,$649 & 0.59 & 16$\,$036 & 0.58 & 0.1 & 224$\,$879 & 0.32 & 102$\,$042 & 0.29 & 0.54 \\ \hline

\multirow{3}{*}{326}& $< 0.25$ & 248$\,$199 & 0.43 & 86$\,$552 & 0.39 & 0.54 & 1$\,$815$\,$778 & 0.12 & 423$\,$255 & 0.08 & 2.26 \\
& $< 0.15$ & 89$\,$234 & 0.53 & 37$\,$328 & 0.49 & 0.23 & 687$\,$466 & 0.2 & 204$\,$036 & 0.15 & 1.09 \\
& $< 0.05$ & 16$\,$469 & 0.62 & 9$\,$050 & 0.61 & 0.06 & 152$\,$973 & 0.33 & 63$\,$194 & 0.3 & 0.34 \\ \hline

\multirow{3}{*}{358}& $< 0.25$ & 188$\,$604 & 0.44 & 83$\,$083 & 0.38 & 0.52 & 1$\,$395$\,$462 & 0.13 & 359$\,$335 & 0.09 & 1.92 \\
& $< 0.15$ & 66$\,$244 & 0.53 & 36$\,$657 & 0.48 & 0.23 & 497$\,$217 & 0.22 & 167$\,$887 & 0.17 & 0.9 \\
& $< 0.05$ & 11$\,$485 & 0.62 & 8$\,$196 & 0.58 & 0.05 & 97$\,$845 & 0.35 & 47$\,$320 & 0.34 & 0.25 \\ \hline

\multirow{3}{*}{389}& $< 0.25$ & 129$\,$532 & 0.46 & 64$\,$698 & 0.41 & 0.4 & 1$\,$172$\,$977 & 0.13 & 323$\,$496 & 0.09 & 1.73 \\
& $< 0.15$ & 39$\,$521 & 0.57 & 24$\,$340 & 0.53 & 0.15 & 397$\,$819 & 0.21 & 136$\,$631 & 0.17 & 0.73 \\
& $< 0.05$ & 5$\,$088 & 0.64 & 3$\,$901 & 0.59 & 0.02 & 73$\,$651 & 0.35 & 37$\,$072 & 0.33 & 0.2 \\ \hline

\multirow{3}{*}{420}& $< 0.25$ & 94$\,$489 & 0.48 & 34$\,$099 & 0.42 & 0.21 & 959$\,$104 & 0.14 & 387$\,$986 & 0.1 & 2.07 \\
& $< 0.15$ & 27$\,$132 & 0.58 & 12$\,$504 & 0.54 & 0.08 & 298$\,$735 & 0.22 & 144$\,$834 & 0.18 & 0.77 \\
& $< 0.05$ & 3$\,$063 & 0.68 & 1$\,$787 & 0.62 & 0.01 & 43$\,$137 & 0.36 & 26$\,$417 & 0.34 & 0.14 \\ \hline

\multirow{3}{*}{452}& $< 0.25$ & 62$\,$988 & 0.51 & 43$\,$781 & 0.47 & 0.27 & 755$\,$731 & 0.15 & 328$\,$130 & 0.12 & 1.75 \\
& $< 0.15$ & 16$\,$231 & 0.59 & 12$\,$851 & 0.55 & 0.08 & 232$\,$231 & 0.24 & 121$\,$850 & 0.21 & 0.65 \\
& $< 0.05$ & 1$\,$566 & 0.67 & 1$\,$381 & 0.62 & 0.01 & 31$\,$731 & 0.38 & 20$\,$606 & 0.36 & 0.11 \\ \hline

\multirow{3}{*}{483}& $< 0.25$ & 53$\,$461 & 0.49 & 53$\,$461 & 0.49 & 0.33 & 600$\,$455 & 0.16 & 600$\,$455 & 0.16 & 3.21 \\
& $< 0.15$ & 12$\,$956 & 0.59 & 12$\,$956 & 0.59 & 0.08 & 165$\,$080 & 0.26 & 165$\,$080 & 0.26 & 0.88 \\
& $< 0.05$ & 1$\,$260 & 0.72 & 1$\,$260 & 0.72 & 0.01 & 20$\,$027 & 0.39 & 20$\,$027 & 0.39 & 0.11 \\ \hline

\end{tabular}
\tablefoot{We also consider the hierarchical framework, by simply adding a subscript $h$ to the counter, e.g. $n_{h, \text{SBLA}}$, and we calculate the incidence rate, $dn/dz$, for these SBLAs. This is presented for the resolved spectra case described in Sect. \ref{sec:variable size}.}
\end{table*}

\section{Variable SBLA Spectral Size in the Non-Hierarchical Framework}\label{appendix:nohierarchy sbla}

\begin{figure*}[h]
    \centering
    \includegraphics[width=0.99\linewidth]{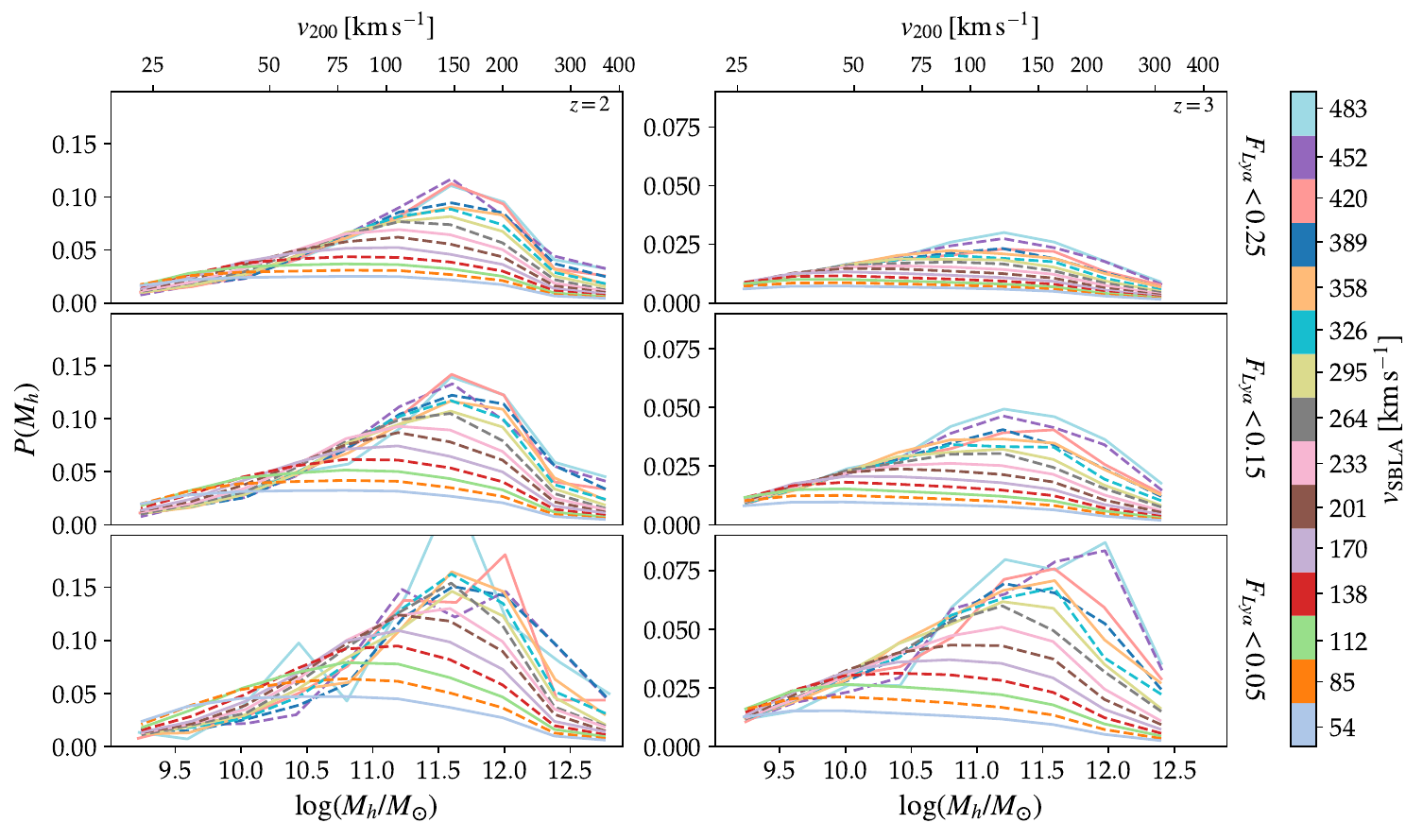}
    \caption{$\pmh$ without using the hierarchical framework. $z=2$ results are on the left panels and $z=3$ results are on the right. All remaining elements are the same as in Fig. \ref{fig:keck psbla cfrac example}.}
    \label{fig:psbla keck random}
\end{figure*}

\begin{figure*}[h]
    \centering
    \includegraphics[width=0.99\linewidth]{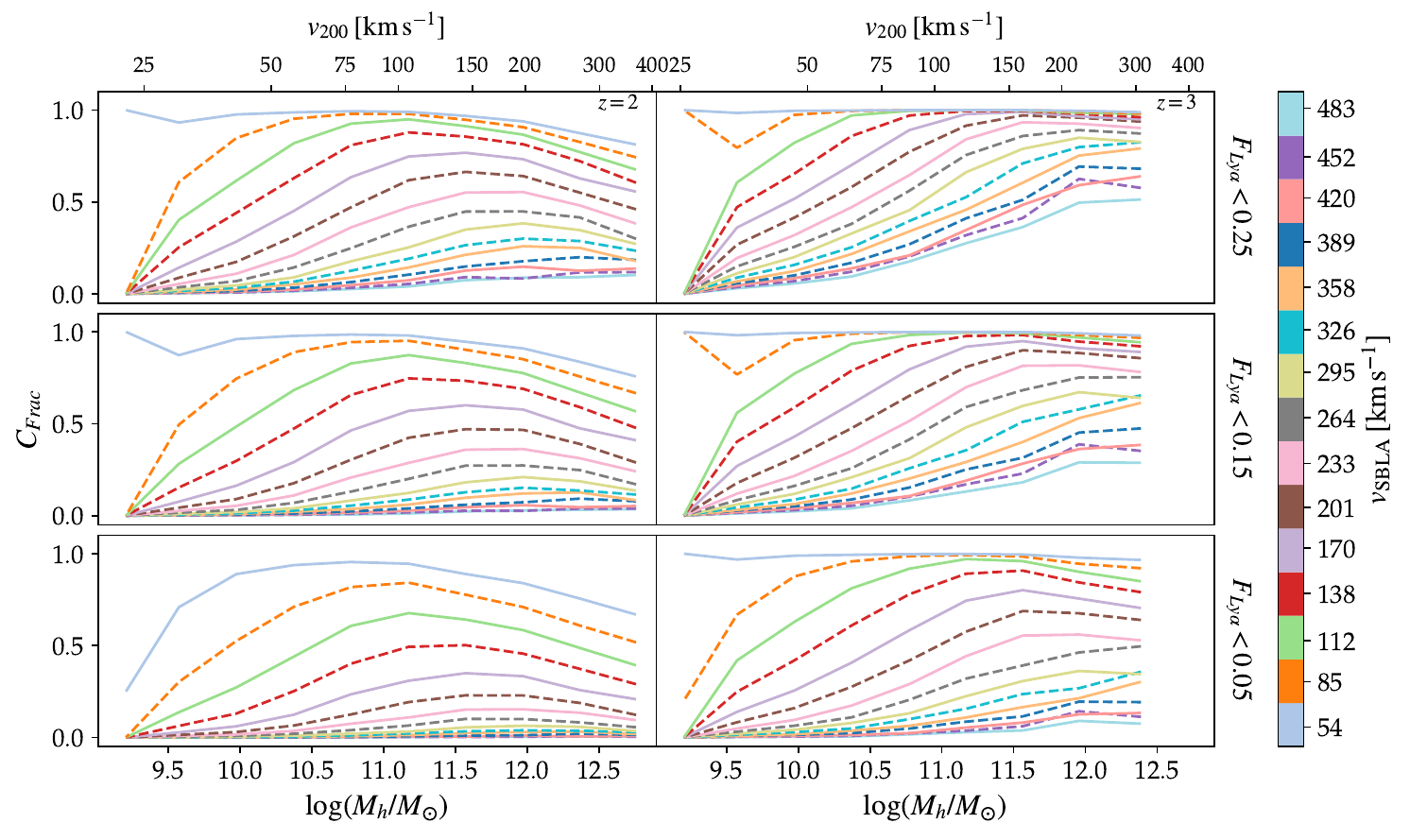}
    \caption{Covering fraction without using the hierarchical framework, divided in mass bins. $z=2$ results are on the left panels and $z=3$ results are on the right. All remaining elements are the same as in Fig. \ref{fig:keck psbla cfrac example}.}
    \label{fig:cov fraction keck random}
\end{figure*}

In this section, we show the results of the $\pmh$ and $\cf$ for the variable SBLAs in the non-hierarchical framework, in Figs \ref{fig:psbla keck random} and \ref{fig:cov fraction keck random}, respectively.

\end{appendix}

\end{document}